\documentclass{article}

\PassOptionsToPackage{square,numbers, compress}{natbib}

\usepackage[preprint]{neurips_2022}

\usepackage[utf8]{inputenc} 
\usepackage[T1]{fontenc}    
\usepackage{hyperref}       
\usepackage{booktabs}       
\usepackage{url}
\usepackage{amsfonts}       
\usepackage{nicefrac}       
\usepackage{microtype}      
\usepackage{xcolor}         

\usepackage{graphicx}
\usepackage{amsmath,amssymb,amsthm}
\usepackage[ruled,vlined,noend]{algorithm2e}
\DeclareMathOperator*{\argmax}{arg\,max}
\DeclareMathOperator*{\argmin}{arg\,min}
\newcommand{\bbE}{\mathbb{E}}
\newcommand{\pp}{\mathcal{P}}

\newcommand{\bi}{\begin{itemize}}
\newcommand{\ei}{\end{itemize}}
\newcommand{\be}{\begin{enumerate}}
\newcommand{\ee}{\end{enumerate}}
\usepackage{subcaption}

\newtheorem{lemma}{Lemma}

\newtheorem{theorem}{Theorem}
\newtheorem{proposition}{Proposition}
\newtheorem{definition}{Definition}

\newcommand\numberthis{\addtocounter{equation}{1}\tag{\theequation}}

\title{Computing the optimal distributionally-robust strategy to commit to}

\author{%
  Sai Mali Ananthanarayanan \\
  Department of Industrial Engineering and Operations Research \\
  and Data Science Institute\\
  Columbia University\\
  New York, NY 10027 \\
  \texttt{sai.mali@columbia.edu} \\
  \And
  Christian Kroer \\
  Department of Industrial Engineering and Operations Research \\
  Columbia University \\
  New York, NY 10027 \\
  \texttt{christian.kroer@columbia.edu} 
}

\begin{document}

\maketitle

\begin{abstract}
The Stackelberg game model, where a leader commits to a strategy and the follower best responds, has found widespread application, particularly to security problems. In the security setting, the goal is for the leader to compute an optimal strategy to commit to, in order to protect some asset. In many of these applications, the parameters of the follower utility model are not known with certainty. Distributionally robust optimization addresses this issue by allowing a distribution over possible model parameters, where this distribution comes from a set of possible distributions. The goal is to maximize the expected utility with respect to the worst-case distribution. We initiate the study of distributionally robust models for computing the optimal strategy to commit to. We consider the case of normal-form games with uncertainty about the follower utility model. Our main theoretical result is to show that a distributionally robust Stackelberg equilibrium always exists across a wide array of uncertainty models. For the case of a finite set of possible follower utility functions we present two algorithms to compute a distributionally robust strong Stackelberg equilibrium (DRSSE) using mathematical programs. Next, in the general case where there is an infinite number of possible follower utility functions and the uncertainty is represented by a Wasserstein ball around a finitely-supported nominal distribution, we give an incremental mixed-integer-programming-based algorithm for computing the optimal distributionally-robust strategy. Experiments substantiate the tractability of our algorithm on a classical Stackelberg game, showing that our approach scales to medium-sized games.
\end{abstract}

\section{Introduction}
\label{sec:intro}
Stackelberg games are a popular game model for settings where one player is able credibly to commit to a strategy, before the other player (or players) get to choose their strategy. The model was originally introduced by von Stackelberg \cite{von1934marktform} in order to analyze competition among firms and first-mover advantage. \cite{Conitzer2006} showed that an optimal strategy for the leader to commit to can be computed in polynomial time. Since then, Stackelberg games have received considerable attention in computational game theory. This has largely been spurred by applications to various security problems such as airport security~\cite{pita2008deployed}, federal air marshal scheduling~\citep{tsai2009iris}, preventing poaching and illegal fishing~\cite{fang2015security}, and several others.

In many applications, the opponent payoffs are not known with certainty. Uncertainty in parameters could occur due to limited scope in observable data, noise or prediction errors. For example, in the case of protecting wildlife from poaching, we would not know the exact utility function of the follower (i.e., the poacher), but only have some model thereof.

One approach for handling uncertainty about payoffs is to assume that each player has a publicly-known distribution over utility functions. This leads to the class of Bayesian Stackelberg games. Computing an optimal strategy for the leader in a Bayesian game setting with a finite set of follower types is NP-hard \cite{Conitzer2006}. Nonetheless, there exist algorithms such as DOBBS \cite{Paruchuri2008PlayingGames} which can solve practical game instances. However, Bayesian games still require an exact specification of the distribution over possible follower types. This again puts strong assumptions on the modeling capacity of the leader who is committing to a strategy based on the supposed distribution.

Another approach to handling uncertainty is through robust optimization. There, the goal is to compute a solution that maximizes utility given the worst-case parameter instantiation~\cite{Ben-tal,Shapiro2020}. In the Stackelberg game context, this is typically interpreted as computing an optimal strategy for the leader to commit to, given that a worst-case follower utility will be selected from an uncertainty set, and then the follower best responds based on this utility function. Robust Stackelberg models and regret-based methods for handling uncertainty about follower payoffs have been studied for security games~\cite{Kiekintveld2010,Kiekintveld2013,Nguyen2014a} as well as \emph{green security games}~\cite{nguyen2015making} and \emph{extensive-form games}~\cite{Kroer2018}.

One issue with robust optimization models is that they can often lead to overly conservative solutions, due to the worst-case nature over a potentially large uncertainty set. In optimization, this can be ameliorated by considering \emph{distributionally-robust optimization} (DRO). 

We briefly describe DRO in a more generic optimization sense, then discuss how to apply DRO models to Stackelberg equilibria. In a standard stochastic optimization problem, we wish to solve an optimization problem of the form
\[
  \min_{x\in X} \underset{\theta\sim \mu}{\mathbb{E}} f(x, \theta),
\]
where $\theta$ is the set of uncertain parameters, and $\mu$ is a distribution over parameters. If the support on $\theta$ is finite then this is comparable to the Bayesian Stackelberg setting, where an exact distribution over parameters is required.

A robust variation of the stochastic optimization problem would replace the expectation with a minimum over $\theta$, where $\theta$ would come from some set that attempts to capture the uncertainty in $\theta$.
In DRO, a middle-ground between stochastic and robust optimization  is struck. In DRO, we assume that there is a distribution over $\theta$, but we do not know it exactly. Instead, that distribution is a worst-case instantiation from some \emph{ambiguity set} of possible distributions. Thus, a generic DRO problem has the form
\[
  \min_{x\in X} \max_{\mu \in \mathcal{D}} \underset{\theta\sim \mu}{\mathbb{E}} f(x, \theta),
\]
where $\mathcal{D}$ is the set of possible distributions over $\theta$.
A recent review of DRO can be found in \citet{Rahimian2019c}. \citet{Shapiro2020} also gives an overview of all these approaches to uncertainty.

Translating the DRO problem into the context of Stackelberg games, $\theta$ is the utility function of the follower, and $\mathcal{D}$ is the set of possible distributions over follower utilities. 

Let us now see how DRO relaxes both the settings of robust Stackelberg games and Bayesian Stackelberg games. First, if the ambiguity set $\mathcal{D}$ is a singleton that contains only the true distribution of follower utility function, then DRO applied to Stackelberg games reduces to a Bayesian Stackelberg game. On the other hand, if $\mathcal{D}$  contains all the Dirac masses on the set of possible follower utilities, then DRO on Stackelberg games reduces to a robust Stackelberg model. Thus, a judicial choice of $\mathcal{D}$ can put DRO between Bayesian games and robust Stackelberg games.  
To the best of our knowledge, there has been no prior work on computing optimal distributionally-robust strategies to commit to.\footnote{\citet{liu2018distributionally} study distributionally-robust Nash and Stackelberg equilibria, but that setting is not about computing the optimal strategy to commit to under distributional uncertainty about the follower utility. Instead, they study a setting where each individual agent employs a DRO approach to their equilibrium behavior. This would not lead to a suitable solution concept when the goal is to compute an optimal leader strategy given uncertainty about the follower utility model, and thus is not suitable for e.g. security games or inspection games with uncertainty about the follower.} Moreover, for the first time, when the set of utility functions could be infinite, we use the Wasserstein metric to construct the ambiguity set for which we have a MIP based algorithm.

\subsection{Our contributions}
We introduce the notion of a distributionally robust Strong Stackelberg Equilibrium (DRSSE) for a normal-form game, allowing us to capture settings where there is some partial knowledge of the follower utility function.
We focus on the simplest case of normal-form Stackelberg games. We prove that a DRSSE is guaranteed to exist for a broad class of ambiguity sets. Our result implies new existence results for some existing robust strong Stackelberg models as well.

We present two direct algorithms to compute DRSSE in the general case, when the set of follower utilities is finite, and with no assumptions on the ambiguity set. First, we show that in this case, it is possible to extend the classical algorithm that enumerates the set of possible best responses for each utility function, and solves a mathematical program for each choice, to the case of DRSSE. In the Bayesian games setting, each mathematical program is an LP, whereas in our setting it is a bilinear saddle-point problem (which can potentially be converted to an LP, depending on the structure of the ambiguity sets). As in the case of Bayesian games, this algorithms requires enumerating an exponentially-large set of mathematical programs, as there are $m^k$ possible choices, where $m$ is the number of follower actions and $k$ is the number of follower utilities. To avoid this exponential search, we introduce binary variables that encode the choice of follower action for each utility function. Then, we show that it is possible to encode the constraints for inducing a given best response for the follower using linear constraints and these binary variables. Putting these things together, we get a mixed-integer program with a bilinear objective.

Next, we allow for an infinitely-large set of possible utility functions $E^f$, and we focus on a representation of the ambiguity sets as a \emph{Wasserstein ball} around some finitely-supported nominal estimate of the probability distribution. Using recent characterizations of such ambiguity sets, we show that in this case we can still use duality theory to arrive at a mixed-integer program, albeit one with a robust optimization flavor which requires repeated MIP solving. Experiments substantiate the tractability of this MIP based algorithm on a classical Stackelberg game, showing that our approach scales to medium-sized games. Implementations of our algorithms and experiments may be found at \url{https://github.com/saimali/DRStackelberg}.
\section{Distributionally Robust Stackelberg Games}
We consider a two player (leader and follower) general-sum game where the leader has a finite set of $n$ actions $A_l$ and the follower has a finite set of $m$ actions $A_f$. Let $\Delta^l, \Delta^f$ denote the set of probability distributions over the leader and follower actions. Let the utilities be $u_l: A_l \times A_f \rightarrow [0,1]$ for the leader and follower utilities $u_f: A_l \times A_f \rightarrow [0,1]$ which come from a compact set $E^f \subseteq [0,1]^{n\times m}$. The arguments inside the utility functions can be mixed strategies in which case, we consider the appropriate weighted sum of utilities, for example, $u_l(x,a_f)  := \sum_{a \in A_l} x_a  u_l(a,a_f)$ when $x \in \Delta^l.$ 
Stackelberg equilibrium is a solution concept for this type of game where we want the leader's strategy to be optimal, assuming the follower will `best respond' knowing the leader's strategy. Given a leader mixed strategy $x \in \Delta^l$, the best response for the follower given a particular utility function $u_f$ is
\begin{equation}\label{eq10}
    BR(x,u_f) = \argmax_{y \in \Delta^f} u_f(x,y)
\end{equation}
One can see that $BR(x,u_f)$ need not be a singleton set. How ties are broken leads to different notions of Stackelberg equilibrium. A few commonly used tie breaking rules are:
\begin{itemize}
    \item \textbf{Strong Stackelberg Equilibrium (SSE),} whenever the follower has multiple best responses, they break ties in favor of the leader. They play $ y^* = \argmax_{y \in BR(x,u_f)} u_l(x,y).$
    \item \textbf{Weak Stackelberg Equilibrium (WSE),} whenever the follower has multiple best responses, they break ties adversarially with respect to the leader. They play $y^* = \argmin_{y \in BR(x,u_f)} u_l(x,y).$
\end{itemize}

Strong Stackelberg equilibrium has been studied by far the most. The SSE assumption is convenient because it usually leads to more tractable solution concepts, and existence is guaranteed, unlike for e.g. WSE.
From a practical standpoint, the assumption that the follower break ties in favor of the leader  is accepted because in most cases the leader can induce the favorable strong equilibrium by selecting a strategy arbitrarily close to equilibrium that makes the follower strictly prefer the desired strategy \citep{Stengel2004} (though see \citet{Guo2018OnGames} for a discussion on how this assumption can fail). More generally, tie-breaking rules rely on precise maximization by the follower and are not always intuitive \cite{Kiekintveld2013}. Therefore robust solutions (like the setting we consider) are useful so as not to depend too much on these rules.

An immediate consequence of the structure of the best response problem is that a pure-strategy best response always exists. To that end, we will only consider pure-strategy best responses throughout the paper.

Now let us consider our robust game model. As mentioned, we assume that there is some set $E^f$ of possible follower utility functions, and the leader does not know which utility function $u_f \in E^f$ the follower will have. Since $u_f(\cdot,\cdot) \in [0,1]$, the set $E^f$ is bounded under most common distance metrics. Strictly speaking we only need $E^f$ to be a Polish space, i.e., a separable complete metric space. However, for ease of readability the reader can think of $E^f$ as endowed with the standard Frobenius norm as a metric. We furthermore assume that there is some (unknown) probability distribution $\mu$ over the set $E^f$. While $\mu$ is unknown, it is assumed to come from some \emph{ambiguity set} $\mathcal{D}_f$, and the goal of the leader is to maximize their worst-case utility over $\mathcal{D}_f$.
Given a leader strategy $x \in \Delta^l$, the worst-case distribution $\mu \in \mathcal{D}_f$ is selected, and overall the leader utility is then 
\begin{equation}
    g(x) := \inf_{\mu \in \mathcal{D}_f}\underset{u_f \sim \mu}{\mathbb{E}} \Bigg[ \  \max_{y \in BR(x,u_f)} u_l(x,y) \Bigg].  \label{eq:leader utility map}
\end{equation} 

The innermost maximization represents the fact that the follower breaks ties in favor of the leader: given $x$ and $u_f$, the follower chooses a best response that is optimal for the leader. The $\inf_\mu \bbE_{u_f}$ part represents the fact that the follower utilities are distributed according to the worst-case distribution $\mu$ chosen from $\mathcal{D}_f$.

An important analytical object will be the inner maximization as a function of $x$ and the follower utility $u_f$:
\begin{equation}\label{def: h function}
h(x,u_f) := \max_{y \in BR(x,u_f)} u_l(x,y).   
\end{equation} 

\begin{lemma}
  The function $h(x,\cdot)$ is upper semicontinuous.
  \label{lem:h(x) usc}
\end{lemma}
\begin{proof}
First, for any $x\in \Delta^l$ and $u_f \in E^f$, the set $BR(x,u_f)$ is non-empty. Let $$y(x,u_f) \in \argmax_{y\in BR(x,u_f)}u_l(x,y)$$ be a mapping from any given pair $(x,u_f)$ to a best response $y(x,u_f)$ that breaks ties in favor of the leader (if there are multiple best responses that break ties in favor of the leader, then we assume that there is a fixed ordering over the finitely-many follower actions, which is used to further break ties).

We wish to show the function $h(x,u_f) = u_l(x,y(x,u_f))$ is upper semicontinuous (u.s.c.) in $x$ for any fixed $u_f$. We use the following definition of upper semicontinuity: for every $x_0$ and $\epsilon > 0$, there exists a neighborhood $N(x_0)$ around $x_0$, such that $h(x,u_f) \leq h(x_0,u_f) + \epsilon$ for all $x \in N(x_0)$.

Now note that for any $y\notin BR(x_0,u_f)$, there exists a neighborhood around $x_0$ such that $y$ is not a best response anywhere in that ball (this follows by the fact that $u_f$ is continuous in $x$). Let $\hat N(x_0)$ be a neighborhood such that this holds for every $y \notin BR(x_0, u_f)$. Also by continuity, we may select a neighborhood $N'(x_0)$ such that $|u_l(x_0,y) - u_l(x,y)| < \epsilon$ for all $x\in N'(x_0)$ and each pure strategy $y$. Now we may select $N(x_0) = \hat N(x_0) \cap N'(x_0)$, which shows that $h(x,\cdot)$ is u.s.c. in $x$.
\end{proof}

Lemma \ref{lem:h(x) usc} helps in proving a crucial property satisfied by $g$: it is upper semicontinuous. This fact will later allow us to easily conclude that equilibria exist in our setting.
\begin{lemma}
  The leader utility function $g(x)$ is upper semicontinuous.
  \label{lem:g usc}
\end{lemma}
\begin{proof}


Define $f(x,\mu) := \mathbb E_{u_f \sim\mu} h(x,u_f),$ and note that $ g(x) = \inf_{\mu \in \mathcal D_f} f(x,\mu).$
  
Upper semicontinuity (u.s.c.) is preserved under pointwise infimum (see e.g. Lemma 2.41 of \cite{aliprantis2006infinite}). Therefore, $g(x) = \inf_{\mu}f(x,\mu)$ is u.s.c. as long as $f(x, \mu)$ is u.s.c. in $x$, for each $\mu$.

Thus it remains to prove that $f(x, \mu)$ is u.s.c. in $x$ for each $\mu$.
By the definition of u.s.c., we want to show that for any sequence $\{x_i\}$ which converges to $x$, $f(x,\mu) \geq \limsup_i f(x_i,\mu)$. Expanding the definition, we have 
\begin{equation*}
    \limsup_i f(x_i,\mu) = \limsup_i \mathbb E_{u_f \sim\mu} h(x,u_f),
\end{equation*} 
by the reverse Fatou Lemma, we then have 
\begin{align*}
\limsup_i f(x_i,\mu) &= \limsup_i \mathbb E_{u_f \sim\mu} h(x_i,u_f) \\
          \leq& \mathbb E_{u_f \sim\mu} \limsup_i h(x_i,u_f).
\end{align*}
To see why the reverse Fatou lemma applies, we need to check two things. First, that there exists a dominating measurable and integrable function. This is clearly true: we take a function which is equal to $1$ on $[0,1]^{n\times m}$ and zero everywhere else. Second, that each $h(x_i,u_f)$ is measurable as a function of $u_f$ when $x_i$ is fixed. To check measurability, first consider that $h(x_i,u_f)$ takes on a finite set of values: $u_l(x_i,a_f)$ for each of the finitely-many $a_f \in A_f$. It suffices to show that for each $a_f$ such that $a_f = y(x_i,u_f)$ for some $u_f$, the nonempty set
$$
U(a_f,x_i) = \{u_f \in [0,1]^{n\times m} : y(x_i,u_f) = a_f\}
$$
is measurable. But this is easily seen, since $U(a_f,x_i)$ is the set of all $u_f$ such that
$u_f(x_i,a_f) \geq u_f(x_i,a_f')$ for all other $a_f'$, and furthermore
$u_f(x_i,a_f) > u_f(x_i,a_f')$ for all $ a_f' $ such that $a_f'$ would be chosen over $a_f$ in case of tied follower utilities.

Finally, applying the fact that $h(x,u_f)$ is u.s.c. gives
\[
\limsup_i f(x_i,\mu) \leq   \mathbb E_{u_f \sim\mu} h(x,u_f)=f(x,\mu).
\] \end{proof}

A distributionally robust Stackelberg solution is a strategy for the leader that maximizes $g(x)$:
\begin{definition}{\textbf{(DRSSS)}}\label{defn1} 
A distributionally robust strong Stackelberg solution (DRSSS) is a mixed strategy $x_l \in \Delta^l$ such that:
\begin{equation}
    x_l \in \argmax_{x \in \Delta^l} \inf_{\mu \in \mathcal{D}_f}\underset{u_f \sim \mu}{\mathbb{E}} \Bigg[ \  \max_{y \in BR(x,u_f)} u_l(x,y) \Bigg]  \label{eq101}
\end{equation} 
\end{definition}

Similar to our definition DRSSS of an optimal leader solution, we can define the corresponding equilibrium concept as well, where we also specify a best response for each utility function of the leader.
\begin{definition}{\textbf{(DRSSE)}}
A strategy tuple $(x,y_{u_f})$ is said to form a distributionally robust strong Stackelberg equilibrium (DRSSE) if $x$ is a DRSSS and $y_{u_f}: E^f \rightarrow A_f$ is a best response mapping which breaks ties in favor of the leader.
\end{definition}

A crucial question is whether DRSSE is guaranteed to exist. One of our main theoretical results is that this is indeed the case. 
The heavy lifting is performed by Lemma~\ref{lem:g usc}, and with Lemma~\ref{lem:g usc} in hand the existence result follows from the extreme value theorem for upper semicontinuous functions.
\begin{theorem}
  \label{thm:drsse exists}
  A DRSSE is guaranteed to exist.
\end{theorem}
\begin{proof}
A pair $(x^*,y(x^*,\cdot))$ is a DRSSE, if $x^* \in \arg\max_{x\in \Delta^l}g(x)$. 
Since the best response set for the follower is always non-empty for a fixed $x^*$ and $u_f$, it suffices to show that $\arg\max_{x\in \Delta^l}g(x)$ is non-empty, or in other words, $g$ attains a maximum over $\Delta^l$. By the extreme value theorem for semicontinuous functions, this is the case if $g$ is u.s.c., which holds by Lemma~\ref{lem:g usc}.
\end{proof}

DRSSE generalizes several existing solution concepts, as the next proposition shows:
\begin{proposition}
\label{prop:drsse generalizes}
    DRSSE generalizes SSE and robust strong Stackelberg equilibrium. 
    DRSSE generalizes Bayesian strong Stackelberg equilibrium when there is only uncertainty about the follower payoff.
\end{proposition}
\begin{proof}
To see why DRSSE generalizes robust Stackelberg, note that we can set the ambiguity set to be the set of point masses on each of the possible utility functions from the robust uncertainty set (as already mentioned in Section~1). 

SSE is the special case of robust SSE where the uncertainty set consists of a single point.

DRSSE generalizes Bayesian strong Stackelberg equilibrium, since a special case of distributional uncertainty is where the ambiguity set consists of a single distribution.
\end{proof}

Combining Theorem~\ref{thm:drsse exists} and Proposition~\ref{prop:drsse generalizes} yields a fairly general existence result: since we can construct robust SSE and Bayesian Stackelberg SSE as special cases, our Theorem~\ref{thm:drsse exists} shows existence for both. This is particularly useful for robust SSE. For example, this implies the first existence result for robust SSE in \emph{extensive-form games}~ \citep{Kroer2018}, where the authors left existence as an open problem.
This holds because an extensive-form game has an equivalent normal-form representation that preserves utilities and the best-response relationship.

\section{Algorithms for Finite Sets of Follower Utilities}
We present two algorithms to compute strategies that form a distributionally robust SSE, in the setting where $E^f$ is a finite set of $k$ possible follower utilities.

Define $\mathcal{Z}$ to be the (finite) set of mappings from follower utility functions to actions. For a particular mapping $z$,  we will use $z_{u_f}$ to denote the action specified for utility function $u_f$ under $z$. Given a leader strategy $x$, there is a at least one $z \in \mathcal{Z}$ that specifies a follower action which is a best response for each possible follower utility. Conversely, given a choice of $z$, we can consider the set $\mathcal{X}_z$ consisting of all strategies for the leader that make $z$ a correct mapping from utility function to best response.

We first show a naive way of computing a DRSSE: we can enumerate all possible $z$, and for each $z$ compute the best possible leader strategy that induces $z$. This constraint on the leader strategy can be captured by a set of linear inequalities, which gives us the following bilinear problem for a fixed $z$:
\begin{gather*}
\text{OPT}_l(z) = \max_{x\in \Delta^l}  \inf_{\mu \in \mathcal{D}_f}\underset{u_f \sim \mu}{\mathbb{E}} \Bigg[ u_l(x,z_{u_f}) \Bigg] \numberthis \label{eq11}   \\
\text{s.t.}\    u_f(x,z_{u_f}) \geq u_f(x,a_f'), \forall \ a_f' \in A_f, \ \forall \ u_f \in E^f 
\end{gather*} 
The set of constraints ensures that for any $u_f$ and $a_f'$, $z_{u_f}$ is among the set of best responses for a leader strategy $x$.
The inner minimization term in the objective represents the fact that even after choosing a best response for each follower utility, the leader still faces the worst-case distribution over those utilities. Since there is a finite set of follower utilities, we can rewrite this as
\begin{equation*}
     \inf_{\mu \in \mathcal{D}_f}\underset{u_f \sim \mu}{\mathbb{E}} \Bigg[ u_l(x,z_{u_f}) \Bigg] =  \inf_{\mu \in \mathcal{D}_f} \Bigg[ \sum_{i=1}^k \mu_{f_i} u_l(x,z_{u_i})  \Bigg],
\end{equation*}
where $z_{u_i}$ is the follower best response for the $i$-th follower utility function.

Now, in order to find the optimal strategy to commit to, we may iterate over all $z \in \mathcal{Z}$ , solve the mathematical program for each, and pick the optimal solution $x^*$ associated to the program with the highest value. Note that if there are multiple best responses to $x^*$, then this approach corresponds to assuming that ties are broken in favor of the leader. Once we have the optimal strategy $x^*$, we may find the associated follower strategy simply by picking the best-response mapping $z$ for which $x^*$ was the solution. Then once an instantiation of the follower utilities is known, the follower plays the corresponding best action taken from this best $z^*$.  This enumeration algorithm shows that DRSSE can be computed in exponential time in terms of $m$ and $k$ (we solve $m^k$ math programs with linear constraints, each in $n$ variables).

In practice, we do not want to enumerate all the exponentially-many possible best response mappings. Instead, we use binary variables to design a mixed-integer non-linear program for branching on the choice of $z$.
Introduce binary variables $\delta_{a_f,u_f}$ for each pair $(a_f,u_f)$ and these variables activate constraints whenever $a_f$ is the BR to $u_f$. For a sufficiently large real $M$,
\begin{align*}
\text{OPT}_l = \max_{x,q} & \inf_{\mu \in \mathcal{D}_f}\underset{u_f \sim \mu}{\mathbb{E}} \bigg[ q_{u_f} \bigg]  \numberthis \label{eq0}\\
\text{s.t.} \ \ & u_f(x,a_f) \geq  u_f(x,a_f') + M(\delta_{a_f,u_f}-1)  \\  
&\ \ \ \ \ \forall \ a_f,a_f' \in A_f, \ \forall \ u_f \in E^f\\
& q_{u_f} \leq u_l(x,a_f) - M(\delta_{a_f,u_f}-1)  \\
&\ \ \ \ \ \forall \ a_f \in A_f, \ \forall \ u_f \in E^f\\
& \sum_{a_f \in A_f} \delta_{a_f,u_f} = 1, \ \forall \ u_f \in E^f\\
& \delta_{a_f,u_f} \in  \{0,1\}\ \  \forall \ a_f \in A_f, \ \forall \ u_f \in E^f \\
& x \in \Delta^l,
 q \in \mathbb{R}^k
\end{align*} \noindent
The first set of constraints relate to picking the best response follower action for each utility function (given a leader strategy $x$). The second set of constraints ensure the leader utility corresponding to the best response follower action above shows up in the innermost term in the objective as $q_{u_f}$.
The third set of constraints guarantees that the first constraint is only activated once for each $u_f$ , i.e., for a given $u_f$, exactly one follower action is assigned as best response. The other constraints specify the domain of $\delta$ (binary variables), $x$ (simplex) and $q$ (real). We prove this math program generates a DRSSE.
\begin{theorem}\label{thm2}
A solution $(x^*,q^*)$ to the mathematical program (\ref{eq0}) forms a DRSSE.
\end{theorem}
\begin{proof}
Since $q$ is being maximized and looking at the constraints, it is clear for any $u_f$, the optimal $q^*_{u_f} = u_l(x,a_f)$ for some $a_f \in A_f$.
It suffices to show that for every $u_f$ and any $x$, 
$$ q^*_{u_f} = u_l(x,a_f) = \max_{y \in BR(x,u_f)} u_l(x,y).$$
If not, there is another $a_f'$ such that for some $u_f$, we have $a_f' \in BR(x,u_f)$ and $$u_l(x,a_f') =  \max_{y \in BR(x,u_f)} u_l(x,y) > u_l(x,a_f) = q^*_{u_f}$$
This is a contradiction to maximality since $$ \inf_{\mu \in \mathcal{D}_f}\underset{u_f \sim \mu}{\mathbb{E}} \Bigg[ u_l(x^*,a_f') \Bigg]  > \underbrace{\inf_{\mu \in \mathcal{D}_f}\underset{u_f \sim \mu}{\mathbb{E}} \Bigg[ q^*_{u_f} \Bigg]}_{OPT_l}$$
Thus the above math program breaks ties in favor of the leader in the strong sense of Stackelberg equilibrium.
\end{proof}

The above mathematical program gives an algorithm to compute a DRSSE. However, it is currently not very practical, due to the objective having an $\inf$ in it. Whether this bilinear objective is easy to handle depends on the form of $\mathcal{D}_f$. Note that here we are heavily exploiting the fact that the set of follower utility functions is finite, in order to encode $z$ using integer variables.

\section{Algorithms for Wasserstein Ambiguity Sets}
We now move on to considering a specific type of ambiguity set $\mathcal{D}_f$ in the case where there is an infinitely-large set of possible utility functions $E^f$, and show that in this case we can still use duality theory to arrive at a mixed-integer program, albeit one with a robust optimization flavor which requires repeated MIP solving.

There are various different ways to deal with ambiguity sets, which should be as small as possible, and contain the true distribution with a good level of certainty\cite{Shapiro2020,Rahimian2019c}. Two major ways are:
\begin{itemize}
    \item \textbf{Define the set using moment constraints}. For example, one can assume moment uncertainty conditions with additional assumptions \cite{Delage2010} to handle robust optimization problems. As discussed in \cite{gao2022}, it has been shown that in many cases these moment based assumptions lets us formulate the problem as a conic quadratic or semi-definite program. However, the moment-based approach is based on the curious assumption that certain conditions on the moments are known exactly but that nothing else about the relevant distribution is known.
    \item\textbf{Distance from a nominal distribution}. A nominal probability distribution $\nu $ in $\mathcal{D}_f$ is given, and $\mathcal{D}_f$ is specified as a set of probability measures which are in some sense close to $\nu$. Popular choices of the statistical distance are $\phi$-divergences (which include Kullback-Leibler divergence and Total Variation distance as special cases), the Prokhorov metric, and Wasserstein distances \cite{gao2022,MohajerinEsfahani2018}. 
 \end{itemize}  
Here we will focus on the second setting, based on distance from a nominal distribution. We leave the question of whether similar results can be obtained for moment-based constraints for future work.
 
Recognizing the fact that the ambiguity set should be chosen judicially for the application at hand, \citet{gao2022} argue that by using the Wasserstein metric the resulting distributions hedged against are more reasonable than those resulting from other popular choices of sets, such as $\phi$-divergence-based sets. Distributionally robust stochastic optimization with Wasserstein distance has been empirically shown to resolve issues with $\phi$-divergences, which do not address how close two points in the support are to each other. The integration involved in the definition of the Wasserstein metric is in a linear form of the joint distribution, whereas typical $\phi$-divergences are nonlinear. For these reasons, we focus on using the Wasserstein metric in our work. 
Our results will build on recent advances in duality theory for dealing with the infinite-dimensional optimization problem over $\mathcal{D}_f$ in the case of Wasserstein distances~\citep{gao2022}.

\subsection{Wasserstein Distance}
Let the set of potential follower utilities $E^f$ be the set of all matrices in $[0,1]^{n\times m}$ specifying a mapping from a strategy pair $a_l,a_f$ to a payoff.
Let $d$ be any distance metric between utility functions such that 
 $E^f$ is a Polish (separable complete metric) space. 
An example metric $d$ would be the Frobenius norm of the difference between the follower payoff matrices: 
$$ d_F(u_{f_i},u_{f_j}) = \left(\sum_{a\in A_l, a'\in A_f} (u_{f_i}(a,a') - u_{f_j}(a,a'))^2\right)^{1/2} $$

Let $\pp(E^f)$ be the set of Borel probability measures on $E^f$, and $\pp_t(E^f)$ its subset with finite $t$-th moment ($t\geq1$). If $\mu,\nu \in \pp_t(E^f)$ (with any metric $d$ on $E^f$), the Wasserstein distance between them is
$$  W_t(\mu,\nu) := \Bigg( \inf_{\gamma \in \Gamma(\mu,\nu)} \int_{E^f \times E^f} d(x,y)^t d\gamma(x,y)   \Bigg)^{1/t}, $$
where $\Gamma(u,v)$ is the collection of all measures with marginals $\mu$ and $\nu$ on the first and second factors respectively. The Wasserstein distance between $\mu,\nu$ is the minimum cost (in terms of $d$) of redistributing mass from $\nu$ to $\mu$. For this reason, it is also called the ``earth mover's distance'' in the computer science literature.
Wasserstein distance is a natural way of comparing two distributions when one is obtained from the other by perturbations. The infimum  is attained if $d$ is lower semicontinuous~\cite{gao2022,zhang2022simple}. 

Using the Wasserstein distance, we define our ambiguity set $\mathcal{D}_f$ as all distributions within a small radius ($\theta$) of a nominal distribution $\nu$:
\begin{equation}\label{eq155}
    \mathcal{D}_f := \{ \mu \in \pp(E^f) \ | \ W_t(\mu,\nu) \leq \theta \}
\end{equation}

The radius $\theta$ controls how far away from the nominal distribution $\nu$ the worst-case distribution can go.
The parameter $\theta$ is also referred to as the \textit{level of robustness}. By adjusting the radius of the ambiguity set, the modeler can thus control the degree of conservatism of the underlying optimization problem. If the radius drops to zero, then the ambiguity set shrinks to a singleton that contains only the nominal distribution, in which case the distributionally robust problem reduces to an ambiguity-free stochastic problem.


Consider the restriction of the inner maximization function $h(x, \cdot) : E^f \to [0,1]$ for a fixed $x$ (see \eqref{def: h function}). We can characterize $h(x, \cdot)$ as a \textit{simple} function, i.e., a measurable function that takes finitely many values $\{u_l(x,y) | y \in A_f\}$. 
Thus it is also $L^1$-measurable.
 
Next we wish to transform the leader utility defined in (\ref{eq:leader utility map}),
into a finite dimensional problem (there could be infinitely many distributions in $D_f$ and we cannot test them all) in the case of a general nominal distribution $\nu \in \pp(E^f)$ and Wasserstein ambiguity set $\mathcal D_f$.
Fix an arbitrary $x$ and let $h(u_f) = h(x,u_f)$. We write this as the \emph{primal} inner problem (akin to the primal problem in \citet{gao2022}), and it is equal to:
\begin{equation}\label{eqn201a}
    \nu_{P} :=   \inf_{\mu \in D^f} \Bigg\{ \int_{E^f} h(u_{f}) \mu(d u_f) : W_t(\mu,\nu)\ \leq \theta \Bigg\},
\end{equation}
Following \citet{gao2022} the dual is then
\begin{equation*}
\nu_D =  \sup_{\lambda \geq 0} \Bigg\{ - \lambda \theta^t + \int_{E^f} \inf_{u_f \in E^f} \bigg[ \lambda d^t(u_f,u_f') + h(u_f) \bigg] \nu(du_f') \Bigg\}
\end{equation*}
It is easily verified that we satisfy all the conditions needed for the strong duality theorem of \citet{gao2022}, and thus we get
$\nu_P = \nu_D < \infty$. The dual is a one dimensional convex minimization problem in $\lambda$, and always admits a minimizer (though in general the infima for each $u_f'$ may or may not have a simple representation). 

\subsection{Nominal distribution with finite support}

We now focus on the setting where the nominal distribution on $E^f$ has finite support. Let the nominal distribution be written as $\nu = \sum_{j=1}^k \nu_j \delta_{\hat{u}_{f_j}}$ for some $\{ \hat{u}_{f_j} \in E^f | j \in [k]\}$.
The finite-support setting is practically important because it occurs when we have received a finite set of observations of follower utilities, and we treat those as an empirical distribution; the Wasserstein ball around this empirical distribution then gives robustness guarantees.

Specializing the primal and dual problems from the general nominal case, for $t \geq 1$ and $\theta>0$ we get
\begin{equation}\label{eqn201finite}
    \nu_{P} :=   \inf_{\mu \in \Delta^k} \Bigg\{ \sum_{i=1}^k \mu_i h(u_{f_i}) : W_t(\mu,\nu)\ \leq \theta \Bigg\}
\end{equation}
\begin{equation*}
\nu_D =  \sup_{\lambda \geq 0} \Bigg\{ -\lambda \theta^t + \sum_{j=1}^k \nu_j \inf_{u_f \in E^f}\bigg[\lambda d^t(u_f,\hat{u}_{f_j}) + h(u_f) \bigg] \Bigg\}
\end{equation*}
The overall problem of computing a DRSSS is then
\begin{align*}
    \text{OPT}_l&(\theta) = \sup_{x \in \Delta^l, \lambda \geq 0} \Bigg\{ -\lambda \theta^t + \sum_{j=1}^k \nu_j w_j : \numberthis \label{eqn1111a}\\
    & w_j \leq \lambda d^t(u_f,\hat{u}_{f_j}) + h(x,u_f) \ , \forall j \in [k],  \ u_f \in E^f \Bigg\} 
\end{align*} 
Computing $\text{OPT}_l$ is not easy since there is an infinite number of constraints due to $E^f$ generally being uncountably large. 

We next propose an incremental MIP-generation approach that addresses this issue.
The key idea behind the MIP is to leverage the fact that $h$ is a simple function in order to represent the constraint for a fixed $u_f$ via several constraints and Boolean variables.
Since there are infinitely-many $u_f$, we start with a small finite set of candidates, and iteratively expand this set.
We proceed in iterations $\tau=1,\ldots$ until convergence.
For each nominal point $j \in [k]$, let $E^f_{\tau,j}$ be the set of utility functions that are considered for point $j$ at iteration $\tau$ of the MIP. Define $E^f_{\tau} := \cup_{j} E^f_{\tau,j}$, whose cardinality is the number of utility functions generated so far. We introduce binary variables $\delta_{a_f,u_f}$ for each pair $(a_f,u_f)$ which denote whether $a_f$ is the chosen best response to $u_f$.
We solve the following MIP at iteration $\tau$,
\begin{align*}
\text{OPT}_l^{\tau} = &\min_{x,\delta,\lambda,w} \bigg\{ \lambda \theta^t -  \sum_{j=1}^k \nu_j w_j \bigg\}  \numberthis \label{mathprog2}\\
\text{s.t.} \ \ & u_f(x,a_f) \geq  u_f(x,a_f') + M(\delta_{a_f,u_f}-1)  \ \ \ \ \ \forall \ a_f,a_f' \in A_f, \ \forall \ u_f \in E_{\tau}^f\\
& w_j \leq (1-\delta_{a_f,u_f})M + \lambda d^t(u_{f},\hat{u}_{f_j}) + u_l(x,a_f) \ \ \ \ \  \forall \ a_f \in A_f, \ \forall \ u_f \in E_{\tau,j}^f, \ \forall j \in [k] \\
& \sum_{a_f \in A_f} \delta_{a_f,u_f} = 1, \ \forall \ u_f \in E_{\tau}^f\\
& x \in \Delta^l, \lambda \geq 0, w \in \mathbb{R}^k \\
& \delta_{a_f,u_f} \in  \{0,1\}\ \  \forall \ a_f \in A_f, \ \forall \ u_f \in E_{\tau}^f
\end{align*} \noindent

The first set of constraints relate to picking the best response follower action for each utility function (given a leader strategy $x$). The second set of constraints is the $w_j$ constraint in $\text{OPT}_l(\theta)$. The third set of constraints guarantees that the first constraint is only activated once for each $u_f$ , i.e., for a given $u_f$, exactly one follower action is assigned as best response. The other constraints specify the domain of $\delta$ (binary variables), $x$ (simplex), $\lambda$ and $w$ (real). 

Using the optimal variables $(x_{\tau},\delta_{\tau},\lambda_{\tau},w_{\tau})$ from solving $\text{OPT}_l^{\tau}$ at iteration $\tau$, we construct a sub-problem that finds new utility functions. For each $j \in [k]$, we define the following subproblem which computes, over the  infinitely-large set $E^f$, the utility function that most violates the second constraint in (\ref{mathprog2})
\begin{align*}
\Gamma(\tau,j) &= \min_{a_f \in A_f} \inf_{u_f \in BR^{-1}(x_{\tau},a_f)} \lambda_{\tau} d^t(u_f,\hat{u}_{f_j}) + u_l(x_{\tau},a_f),  \numberthis \label{subproblem}  \\
\end{align*}
Let $u_f(\tau,j)$ be the utility function in $E^f$ that gives us $\Gamma(\tau,j)$. 
Next, we then compute the most-violated of the  $k$ nominal points
\begin{align*}
\Gamma(\tau) &:= \min_{j \in [k]} \{ \Gamma(\tau,j) - w_{\tau_{j}} \} \numberthis \label{Gamma Tau}
\end{align*} \noindent
Since we can enumerate over $a_f$, the only hard part of solving these subproblems is to resolve the inner $\inf$ term, calculated over $BR^{-1}(x,a_f)$, which may be difficult depending on the structure of $E^f$.
If we have an oracle that gives us the inner $\inf$ term for a fixed $a_f$, we can solve the subproblem in $m$ oracle calls.

Next we show that when $E^f$ is described by linear constraints, the inner $\inf$ problem can be solved via convex minimization as long as the distance metric $d$ is nice (e.g. for the $\ell_1$ or $\ell_2$ distance).
For a fixed $a_f \in A_f$, the inner problem can be written as,
\begin{align*}
\inf_{u_f} \ \  &  \lambda_{\tau} d^t(u_f,\hat{u}_{f_j}) + u_l(x_{\tau},a_f)  \numberthis \label{subproblem structure: appendix} \\
\text{s.t.}\ \ & \sum_{i=1}^n x_{\tau,i} \bigg[ u_f(a_i,a_f) - u_f(a_i,a'_{f}) \bigg]  > 0  \ \ \forall a'_f \in B_{x_{\tau}}(a_f), \\
&\sum_{i=1}^n x_{\tau,i} \bigg[ u_f(a_i,a_f) - u_f(a_i,a'_{f}) \bigg]  \geq 0  \ \  \forall \ a_f' \in A_f \setminus B_{x_{\tau}}(a_f), \\
& u_f \in E^f,
\end{align*} \noindent
where $B_{x}(a) := \{a_f' \in A_f : u_l(x,a'_f) > u_l(x,a) \}$. Each element $a_f'$ of $B_x(a) \subset A_f$ (known deterministically for a given $x$ and $a$) has leader utility bigger than $a$. Because we wish to enforce strong Stackelberg tie-breaking within the best response set, we must ensure that any $a_f' \in B_{x_{\tau}}(a_f)$ is not picked instead of $a_f$, and hence, the first constraint of (\ref{subproblem structure: appendix}) picks an $u_f$ such that $a_f$ is strictly a best response to a given leader strategy $x_{\tau}$ compared to $a_f' \in B_{x_{\tau}}(a_f)$, and the second constraint allows for $a_f' \notin B_{x_{\tau}}(a_f)$ to also be best responses along with $a_f$.
In practice, strict inequalities are handled by industry-grade solver, typically by enforcing a very small epsilon gap.

Thus for a nice distance metric $d$ (such as $\ell_1$ or $\ell_2$ distance), the subproblem (\ref{subproblem structure: appendix}) is a convex minimization problem with linear constraints. 

We can now summarize the algorithm to solve (\ref{mathprog2}). We start with each $E_{1,j}^f$ to be just the nominal functions $\hat{u}_{f_j}$, and at each iteration $\tau$, solve the program $\text{OPT}_l^{\tau}$. Use the optimal solution to solve the subproblem (\ref{subproblem}) for each $a_f \in A_f$ to add more utility functions to $E_{\tau+1,j}^f$, until there is an iteration where none can be added (thus $\Gamma(\tau) \geq 0$).
\begin{algorithm}
\DontPrintSemicolon
\SetAlgoLined
$\tau \leftarrow  1$ \tcp{iteration counter}
$E_{\tau,j}^f \leftarrow  \{ \hat{u}_{f_j}\} \ \ \forall j \in [k]$ 

$\Gamma \leftarrow  -1$

\While{$\Gamma < 0$}{
    Solve the MIP (\ref{mathprog2}) for $\text{OPT}^{\tau}_l$ and obtain 
    the solution $(x_{\tau},\delta_{\tau},\lambda_{\tau},w_{\tau})$ \;

    \For{$j = 1,\ldots,k$}{
    Compute $\Gamma(\tau,j)$ as described in (\ref{subproblem})\;
    
    \If{$\Gamma(\tau,j) < w_{\tau_j}$}{
    $E_{\tau+1,j}^f \leftarrow  E_{\tau,j}^f \cup \{ u_f(\tau,j)$ \}
    }
    }
    Update $\Gamma \leftarrow  \Gamma(\tau)$ as defined in (\ref{Gamma Tau}) \;
    Update $\tau \leftarrow \tau + 1$ \;
    }
Output $(x^{*},\delta^{*},\lambda^{*},w^{*}) = (x_{\tau-1},\delta_{\tau-1},\lambda_{\tau-1},w_{\tau-1})$
 \caption{MIP Algorithm to compute DRSSS}
  \label{alg:DRSSE MIP}
\end{algorithm} \noindent
Thus Algorithm \ref{alg:DRSSE MIP} can be used to compute the optimal leader strategy $x^*$ and DRSSS.

\section{Experiments}
We now test the scalability of the MIP Algorithm \ref{alg:DRSSE MIP} (henceforth referred to as \textit{DR MIP}). We do not consider any baselines in our general setting with an infinite set of utility functions. A discussion of Wasserstein ambiguity sets with finite $E^f$ ,along with suitable baselines is provided in the Appendix.

We present the experimental performance based on a classic Stackelberg game from GAMUT \cite{nudelman2004run}, and a synthetic data set in the Appendix. We vary different game parameters to investigate the scalability. All experiments are timed out at 1000 seconds, and run times reported as a function of the parameter being varied. All experiments were conducted using Gurobi $9$ to solve MIPs and LPs (default internal parameters), on a Macintosh with $2.4$ GHz Quad-Core intel Core $i5$ processor. Run times are reported in seconds. We consider a typical data-driven setting, where the nominal distribution $\nu$ is simply the discrete uniform distribution with weight $\frac{1}{k}$ on each of the $k$ empirical observations. We set the Wasserstein radius $\theta=0.1$, exponent $t=2$ and a tight choice of $M=2$ for the MIP (\ref{mathprog2}).

\begin{figure*}[!ht]
\begin{subfigure}{.45\textwidth}
\centering
  \includegraphics[width=\textwidth]{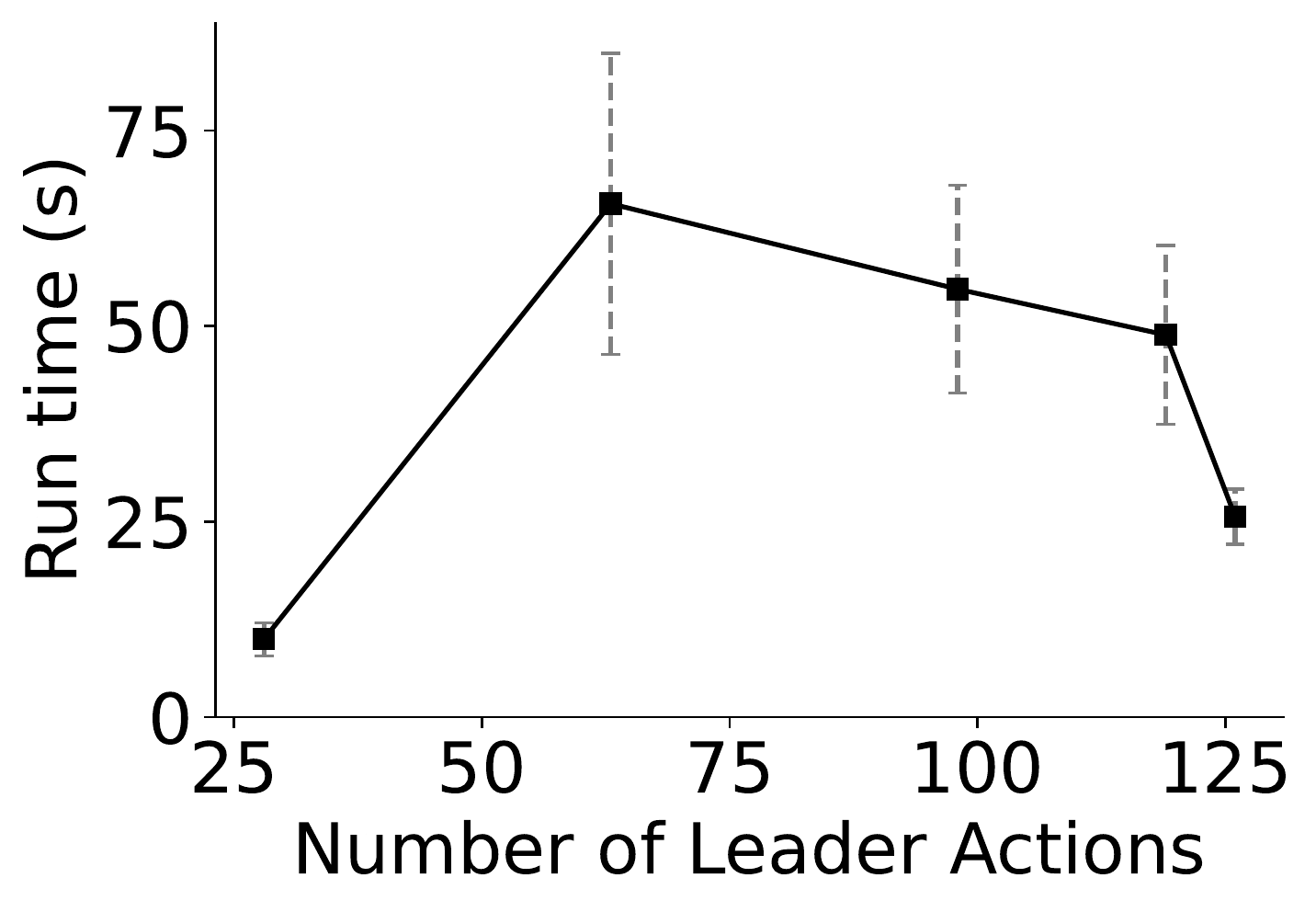}
  \caption{}
  \label{fig:2a}
\end{subfigure} 
\begin{subfigure}{.45\textwidth}
\centering
   \includegraphics[width=\textwidth]{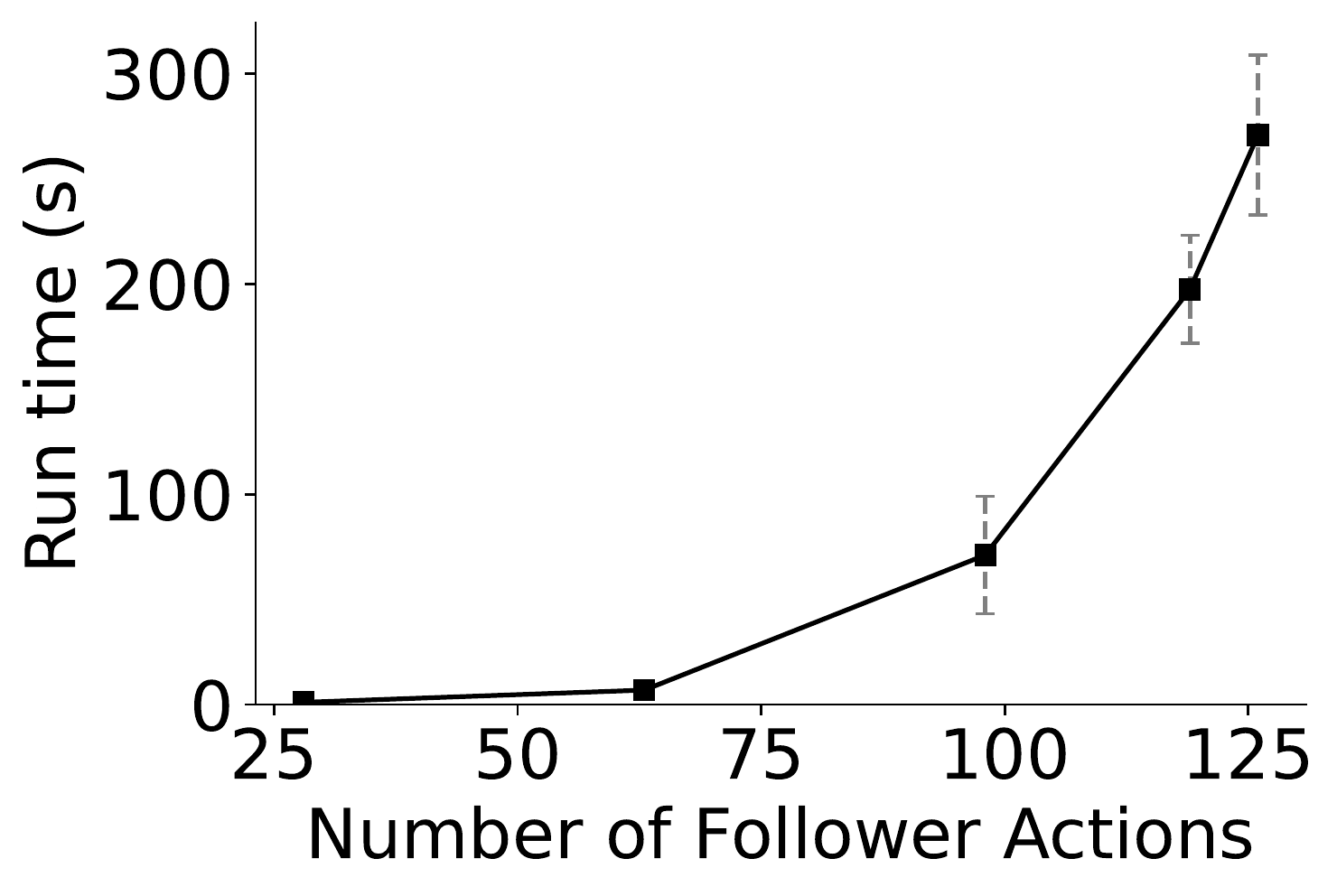}
  \caption{}
  \label{fig:2b}
\end{subfigure} \\
\begin{subfigure}{.45\textwidth}
\centering
   \includegraphics[width=\textwidth]{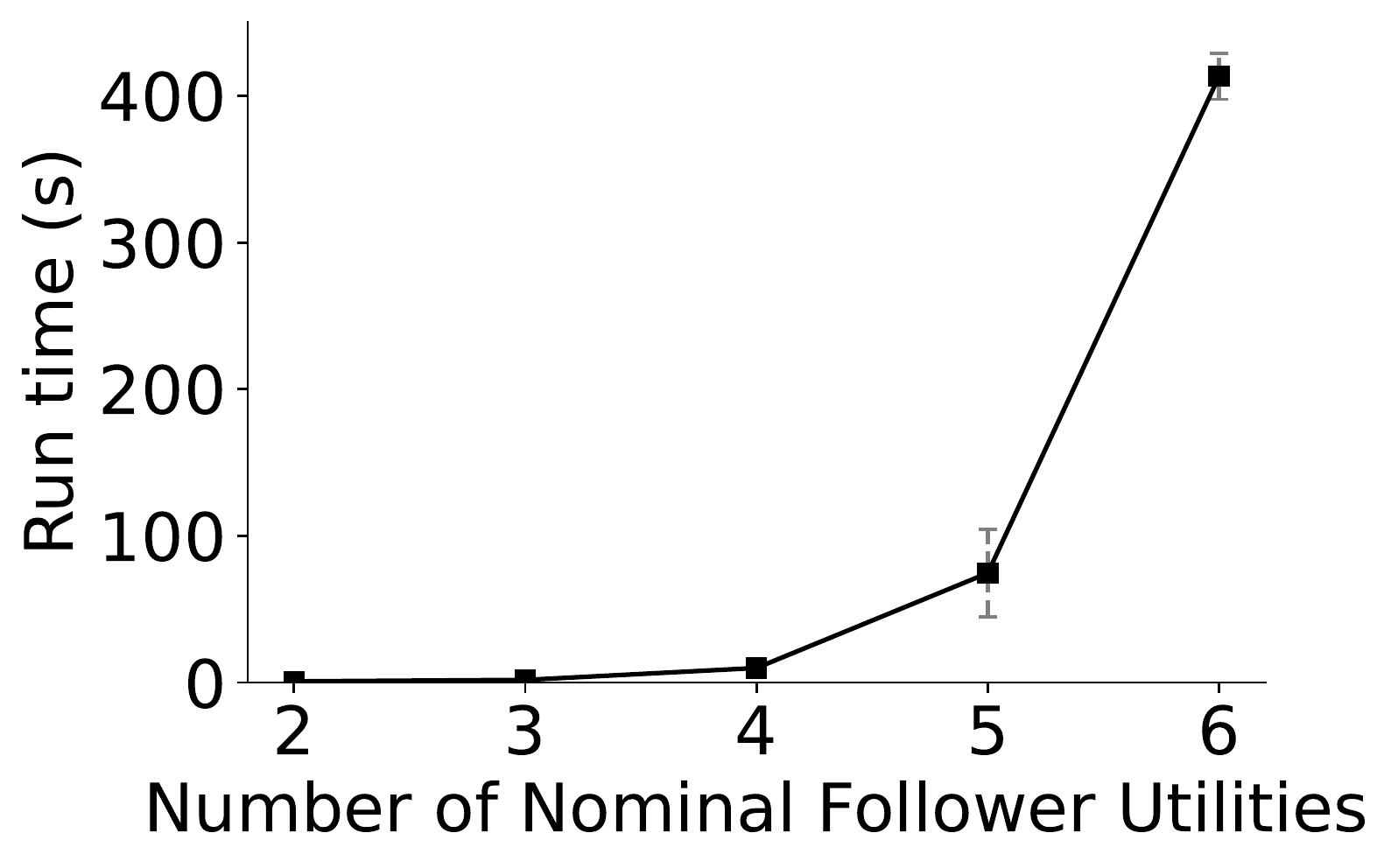}
  \caption{}
  \label{fig:2c}
\end{subfigure}
\begin{subfigure}{.45\textwidth}
\centering
   \includegraphics[width=\textwidth]{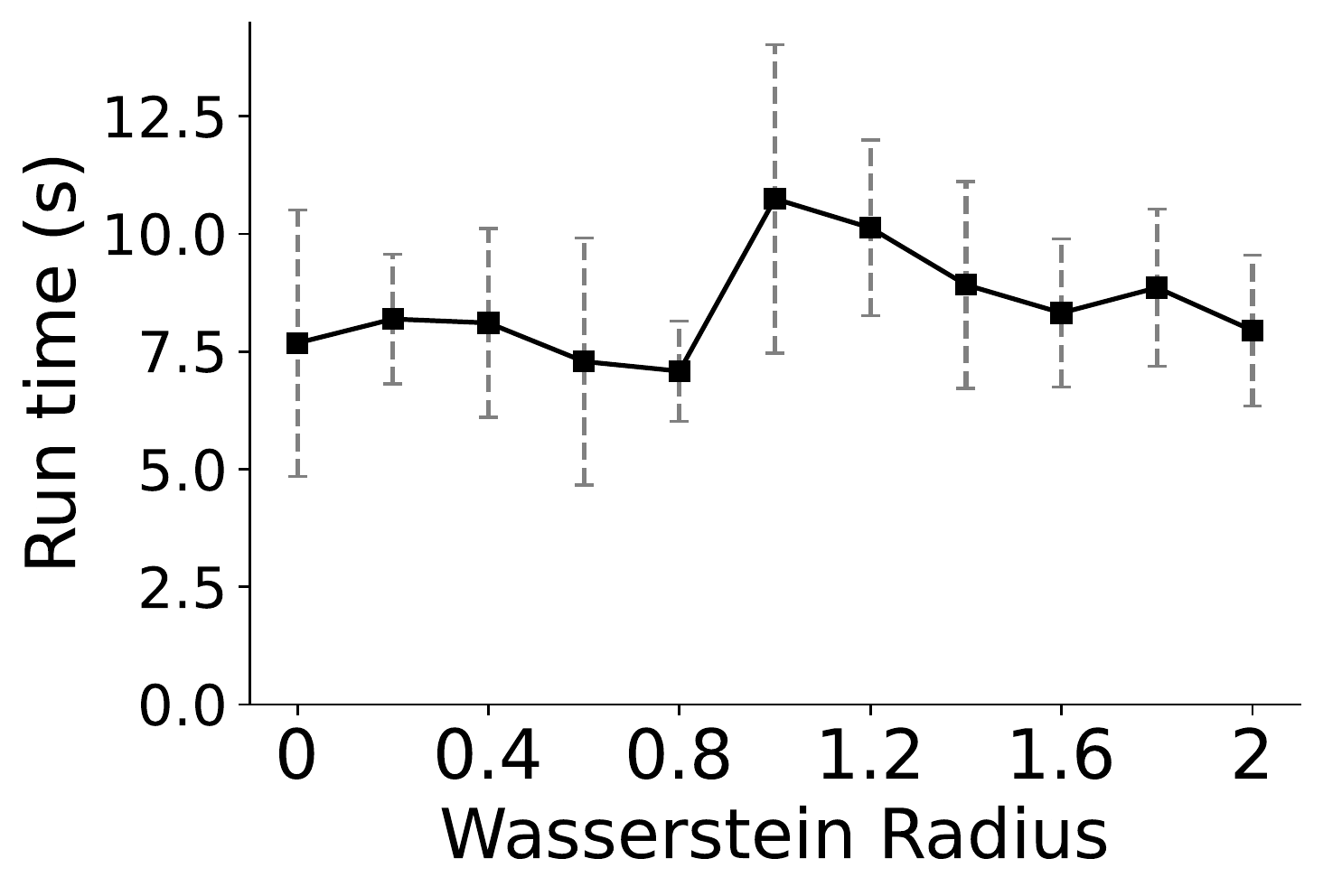}
  \caption{}
  \label{fig:2d}
\end{subfigure}
\caption{Performance of the DR MIP based Algorithm 1 on the Simple Inspection Game, averaged over $10$ simulations with standard deviations displayed. \textbf{(a)} Runtime vs number of leader actions ($n$) with $s=7, q=2$ and $k=4$. \textbf{(b)} Runtime vs number of follower actions ($m$) with $s=7, p=5$ and $k=2$. \textbf{(c)} Runtime vs number of nominal follower functions ($k$) with $s=7, p=2$ and $q=2$. \textbf{(d)} Runtime vs Wasserstein radius ($\theta$) with $s=7, p=2, q=2$ and $k=4$. }
\label{fig:2}
\end{figure*} \noindent

\paragraph{Results on Inspection Game}
The Inspection Game~\cite{avenhaus2002inspection} is a classic Stackelberg setting where an inspector tries to deter an inspectee from cheating. In the Simple Inspection Game setting in GAMUT, there is a set $S$ of size $0 < s \leq 8$. The inspector (the leader) chooses a subset of $S$ of size at most $p \ (0 < p \leq s)$. Hence the size of their action space is $n = \binom{s}{1} + \ldots + \binom{s}{p}$. Similarly the inspectee (the follower) chooses a subset of size at most $q$. Hence the size of their action space is $m = \binom{s}{1} + \ldots + \binom{s}{q}$. If there is no intersection in the chosen sets, the leader receives a payoff $-\alpha$ and the follower receives a payoff $\alpha$ for some $\alpha>0$. Otherwise both players get zero payoff. The structure of $E^f$ is helpful in tractably computing the subproblem (\ref{subproblem}) for the Inspection game. In fact, we can express the objective of the inner subproblem (\ref{subproblem structure: appendix}) as a quadratic function of two variables for a $\ell_2$ distance metric such as Frobenius norm $d_F$. Recall the definition,
$$ d_F(u_{f_i},u_{f_j}) = \left(\sum_{a\in A_l, a'\in A_f} (u_{f_i}(a,a') - u_{f_j}(a,a'))^2\right)^{1/2} $$
The utility function of the inspection game has only two variables, the payoff $\alpha$ when the sets intersect and the payoff $\beta$ when the sets do not intersect. For a given inspection game size, the positions of $\alpha$ in any follower utility matrix $u_f$ is the same (resp. for $\beta$). And in the $nm$ entries of the matrix, a fixed number (say $c$) of the entries have value $\alpha$, and the rest have value $\beta$. 
Therefore the distance function in the objective of (\ref{subproblem structure: appendix}) can be written as,
\begin{align*}
    d_F(u_{f},\hat{u}_{f_j}) &= \left(\sum_{a\in A_l, a'\in A_f} (u_{f}(a,a') - \hat{u}_{f_j}(a,a'))^2\right)^{1/2} \\
    &= \left(c (\alpha_{u_f} - \alpha_{\hat{u}_f})^2 + (nm-c) (\beta_{u_f} - \beta_{\hat{u}_f})^2 \right)^{1/2}
\end{align*}
In our experiments on Inspection game, we choose $t=2$, and hence the the inner subproblem (\ref{subproblem structure: appendix}) is a quadratic program in two variables $\alpha$ and $\beta$ denoting the payoffs of the Inspection follower utility $u_f$. The objective is simply
\begin{align*}
  &\inf_{\alpha,\beta} \bigg\{ \lambda_{\tau} c (\alpha - \alpha_{\hat{u}_f})^2 + \lambda_{\tau} (nm-c) (\beta - \beta_{\hat{u}_f})^2 \bigg\},
\end{align*}
and we can write the linear constraints similarly as well.

In the following experiment we normalize the payoffs to lie in $[0,1]$ and set leader payoffs to lie in $\{0,0.5\}$ instead of $\{-\alpha,0\}$. For $k$ different nominal follower utility functions, we set follower payoffs to be random variables that are uniformly distributed in $[0.3,0.6)$ instead of $0$ and uniformly distributed in $[0.7,1)$ instead of $\alpha$.

In Figure \ref{fig:2a}, for $s=7,q=2$, and $k=4$, we vary the number of leader actions (by varying the maximum size $p$ of the leader set). Since the leader wishes for an intersection to happen, picking a larger set is good. As $p$ increases, the size of the leader action space increases and the runtime of the DR MIP goes up. However for $p =5$ and $p=6$, larger sets are readily available and the DR MIP converges faster though the leader action space is large. 

In Figure \ref{fig:2b}, for $s=7,p=5$, and $k=2$, we vary the number of follower actions (by varying the maximum size $q$ of the follower set). Recall that the number of binary variables $\delta$ in the iterative MIP scales with the size of follower set, and hence the number of follower actions again has a moderate impact on scalability of the DR MIP.

In Figure \ref{fig:2c}, for $s=7,p=q=2$, we vary the number of nominal utility functions which has an exponential impact on the scalability of the DR MIP, thus having access to a a higher of empirical distributions slows down the algorithm. Recall again that the number of binary variables is also dependent on $k$ at each iteration of the algorithm.

In Figure \ref{fig:2d}, for $s=7,p=2, q=2$, and $k=4$, we vary the Wasserstein radius $\theta$ from $0$ to $2$. There is little impact on the running time by considering an increase in this parameter. 

\section{Conclusion}
In this work, we initiated the study of computing optimally distributionally robust strategies to commit to. 
We formalized the notion of a distributionally robust strong Stackelberg solution for the leader, and showed that these are guaranteed to exist in a wide number of settings. We presented two algorithms for computing a DRSSE using mathematical programs for any ambiguity set. One algorithm has only continuous variables and the other has mixed integer variables, and the constraints are linear. When the uncertainty is represented by Wasserstein uncertainty, we showed that the above programs can be solved with an incremental mixed-integer linear program. We performed computational experiments on the MIP in terms of different parameters on a classical Stackelberg game where the structure of the set of utility functions can be exploited to compute subproblem tractably. In the Inspection game, the MIP based algorithm scaled well for medium-sized games. We found that the runtime impact of the size of the leader action set is low, the number of nominal utility functions has high (exponential) impact, and the size of the follower action space has moderate impact. 

One avenue for future work is to study ambiguity sets that are described by moment uncertainty conditions. These type of assumptions lead to conic quadratic or semi-definite programs in many settings \cite{gao2022}. For DRSSE, it would be interesting to see if it is possible to derive a mixed-integer conic program based on these results. Another promising avenue would be to take our general results on distributionally robust Stackelberg equilibria and interpret them for popular applications of Stackelberg games like security games, where one could potentially exploit problem structure in order to get more scalable algorithms. 

\typeout{}
\bibliographystyle{plainnat}
\bibliography{references.bib}

\appendix
\setcounter{equation}{0}
\renewcommand\theequation{A.\arabic{equation}}

\section{Special case: Wasserstein ambiguity sets with a finite set of utility functions}
Let us analyze the case where the set of follower utility functions $E^f = \{u_{f_1},\ldots,u_{f_k}\}$ is finite.
Since the support $E^f$ of the distributions is finite, let the nominal distribution be written as $\nu = \sum_{j=1}^k \nu_j \delta_{u_{f_j}} $ and any other distribution as $\mu = \sum_{i=1}^k \mu_i \delta_{u_{f_i}} $, where $\delta_{u_f}$ denotes the unit mass on $u_f$. Since $E^f$ is finite, the integral in the definition of Wasserstein metric simplifies to a summation and can be written as,
\begin{align*}
W_t^t(\mu,\nu) &:= \min_{\gamma_{ij} \geq 0} \Bigg\{ \sum_{i=1}^k \sum_{j=1}^k d^t(u_{f_i},u_{f_j}) \gamma_{ij} \\
 &\ \ \ \ \ : \sum_{j=1}^k \gamma_{ij} = \mu_i \ \forall \ i, \sum_{i=1}^k \gamma_{ij} = \nu_j  \ \forall j \Bigg\} \numberthis \label{deff1}
\end{align*}

We also write the dual of the Wasserstein metric as,
\begin{align*}
W_t^t(\mu,\nu) &:= \max_{(r,s) \in \mathbb{R}^k \times \mathbb{R}^k} \Bigg\{ r^T \mu + s^T \nu  \\ 
 &\ \ \ \ \ : r_i + s_j \leq d^t(u_{f_i},u_{f_j}) \ \forall \ i, j \Bigg\} \numberthis \label{deff2}
\end{align*}

Applying Theorem 1 from \cite{gao2022}, we get strong duality and the overall problem of computing a DRSSS is
\begin{align*}
    \text{OPT}_l &= \min_{\substack{z \in A_f^k \\ x \in \mathcal{X}_z}}  \inf_{\lambda \geq 0} \Bigg\{ \lambda \theta^t -  \sum_{j=1}^k \nu_j w_j \\
    &\ \ \ \ : w_j \leq \lambda d^t(u_{f_i},u_{f_j}) + u_l(x,z_{u_{f_i}}) \ \forall i, j \Bigg\}, \numberthis \label{dual: finite case}
\end{align*}
where the set $\mathcal{X}_z$ is defined as follows: if we fix any $z \in \mathcal{Z}$, the set of feasible leader actions is restricted to the set
\begin{align*}
    \mathcal{X}_z &= \{ x \in \Delta^l \ : u_{f_j}(x,z_{j}) \geq u_{f_j}(x,a_f) \\
    &\ \ \ \ \ \ \ \   \forall \ 1 \leq j \leq k, \ \ \forall \ a_f \in A_f \}
\end{align*}  
Notice in the program for OPT$_l$ that while $z$ is from a more general space, once a $z$ is fixed, we have to pick $x$ from the set $\mathcal{X}_z$, therefore the objective is not easy to compute. To deal with the constraints in $\mathcal{X}_z$,
we define boolean variables $\delta_{a_f,u_f}$ for each pair $(a_f,u_f)$ and use these variables to activate the constraints in the definition of $\mathcal{X}_z$, as well as the constraints involving $w_j$'s. For a sufficiently large $M$,
\begin{align*}
\text{OPT}_l = &\min_{x,y,\lambda,w} \bigg\{ \lambda \theta^t -  \sum_{j=1}^k \nu_j w_j \bigg\}  \numberthis \label{mathprog:appendix}\\
\text{s.t.} \ \ & u_f(x,a_f) \geq  u_f(x,a_f') + M(\delta_{a_f,u_f}-1)  \\  
&\ \ \ \ \ \forall \ a_f,a_f' \in A_f, \ \forall \ u_f \in E^f\\
& w_j \leq (1-\delta_{a_f,u_f})M + \lambda d^t(u_{f},u_{f_j}) + u_l(x,a_f) \\
&\ \ \ \ \  \forall \ a_f \in A_f, \ \forall \ u_f \in E^f, \ \forall j \\
& \sum_{a_f \in A_f} \delta_{a_f,u_f} = 1, \ \forall \ u_f \in E^f\\
& x \in \Delta^l, \lambda \geq 0, w \in \mathbb{R}^k \\
& \delta_{a_f,u_f} \in  \{0,1\}\ \  \forall \ a_f \in A_f, \ \forall \ u_f \in E^f
\end{align*} \noindent
We obtain a MIP similar to iterative MIP (11) in the paper, but due to the finiteness of $E^f$, we can solve (\ref{mathprog:appendix}) directly to get the leader strategy and compute DRSSS. The MIP has $n + k + 1 $ continuous variables, $mk$ binary variables, and $m^2k + m k^2 + k + 2 = O(mk (m+k))$ linear constraints. 

\subsection{Baselines}
We illustrate experiments on the case where $E^f$ is finite on two different games- complementing the Inspection Game considered in the main body, we performing experiments on another classical Stackelberg game in GAMUT called Cournot Duopoly, as well as the synthetic dataset of all possible random matrices in $[0,1]$, the most general ground set of the utility functions.

We compare the DR MIP (\ref{mathprog:appendix}) in both games on two baselines, explained below. 

(a) the first baseline is an an enumeration approach where we enumerate all possible instances of $z$ in (\ref{dual: finite case})  and for each $z$, we solve the LP and take the best result; Define the LP given a $z$ (written as boolean variables $\delta$),
\begin{align*}
&\text{OPT-LP}(\delta) = \min_{x,\lambda,w} \bigg\{ \lambda \theta^t -  \sum_{j=1}^k \nu_j w_j \bigg\}  \numberthis \label{mathprogLP: appendix}\\
\text{s.t.} \ \ & u_f(x,a_f) \geq  u_f(x,a_f') + M(\delta_{a_f,u_f}-1)  \\  
&\ \ \ \ \ \forall \ a_f,a_f' \in A_f, \ \forall \ u_f \in E^f\\
& w_j \leq (1-\delta_{a_f,u_f})M + \lambda d^t(u_{f},u_{f_j}) + u_l(x,a_f) \\
&\ \ \ \ \  \forall \ a_f \in A_f, \ \forall \ u_f \in E^f, \ \forall j \\
& x \in \Delta^l, \lambda \geq 0, w \in \mathbb{R}^k 
\end{align*} \noindent
If the set of all possible instances of $\delta$ is given by $$\mathcal{Q} = \bigg\{ \delta \in \{0,1\}^{m \times k} :  \sum_{i} \delta_{ij} = 1, \ \forall \ j   \bigg\},$$ then the enumeration approach is computing the best result among $m^k$ LPs,
$$ \text{OPT}_l = \max_{\delta \in \mathcal{Q} }  \text{OPT-LP}(\delta).$$

(b) the second baseline is the Bayesian Stackelberg MIP which is non-robust, where the ambiguity set is a singleton set with only a nominal distribution.
\begin{align*}
&\text{OPT-Bayesian}(\nu) = \max_{x,\delta,w}  \sum_{j=1}^k \nu_j w_j  \numberthis \label{mathprog: Bayesian appendix}\\
\text{s.t.} \ \ & u_f(x,a_f) \geq  u_f(x,a_f') + M(\delta_{a_f,u_f}-1)  \\  
&\ \ \ \ \ \forall \ a_f,a_f' \in A_f, \ \forall \ u_f \in E^f\\
& w_j \leq (1-\delta_{a_f,u_f})M +  u_l(x,a_f) \\
&\ \ \ \ \  \forall \ a_f \in A_f, \ \forall \ u_f \in E^f, \ \forall j \\
& \sum_{a_f \in A_f} \delta_{a_f,u_f} = 1, \ \forall \ u_f \in E^f\\
& x \in \Delta^l, w \in \mathbb{R}^k \\
& \delta_{a_f,u_f} \in  \{0,1\}\ \  \forall \ a_f \in A_f, \ \forall \ u_f \in E^f. \\
\end{align*} \noindent

We now test the scalability of the distributionally robust MIP (\ref{mathprog:appendix}) and use the two baselines (\ref{mathprogLP: appendix}) and (\ref{mathprog: Bayesian appendix}). The nominal distribution $\nu$ is generated as a random probability vector, and the rest of the experimental setup is the same as the main body. 
\subsection{Results on Cournot Duopoly Game}
The Cournot duopoly \cite{cournot1897researches} is a game that models two rival firms choosing the quantity of competing goods to produce at the same time. We consider the Stackelberg equlibrium setting of this game in GAMUT. Given an inverse demand function $P(\cdot)$ and increasing cost functions $C_i$ for player $i$, the utility for player $i$ given player actions $(y_1,y_2)$ is $u_i(y_1,y_2) = P(y_1+y_2) \times y_i - C_i(y_i)$. With the notation $randint(a,b)$ to denote a random integer between $a$ and $b$, in the following experiment we set $P(x) = 75 - randint(1,10)x, C_1(y) = randint(10,40) + randint(10,20)y$ and $ C_2(y) = randint(2,20) + randint(1,5)y $ to compute the utilities and normalize them to lie in $[0,1]$. The number of actions is equal for both players in the game ($n=m)$.

\renewcommand{\thefigure}{A\arabic{figure}}
\setcounter{figure}{0} 
\begin{figure*}[!ht]
\begin{subfigure}{.45\textwidth}
\centering
  \includegraphics[width=\textwidth]{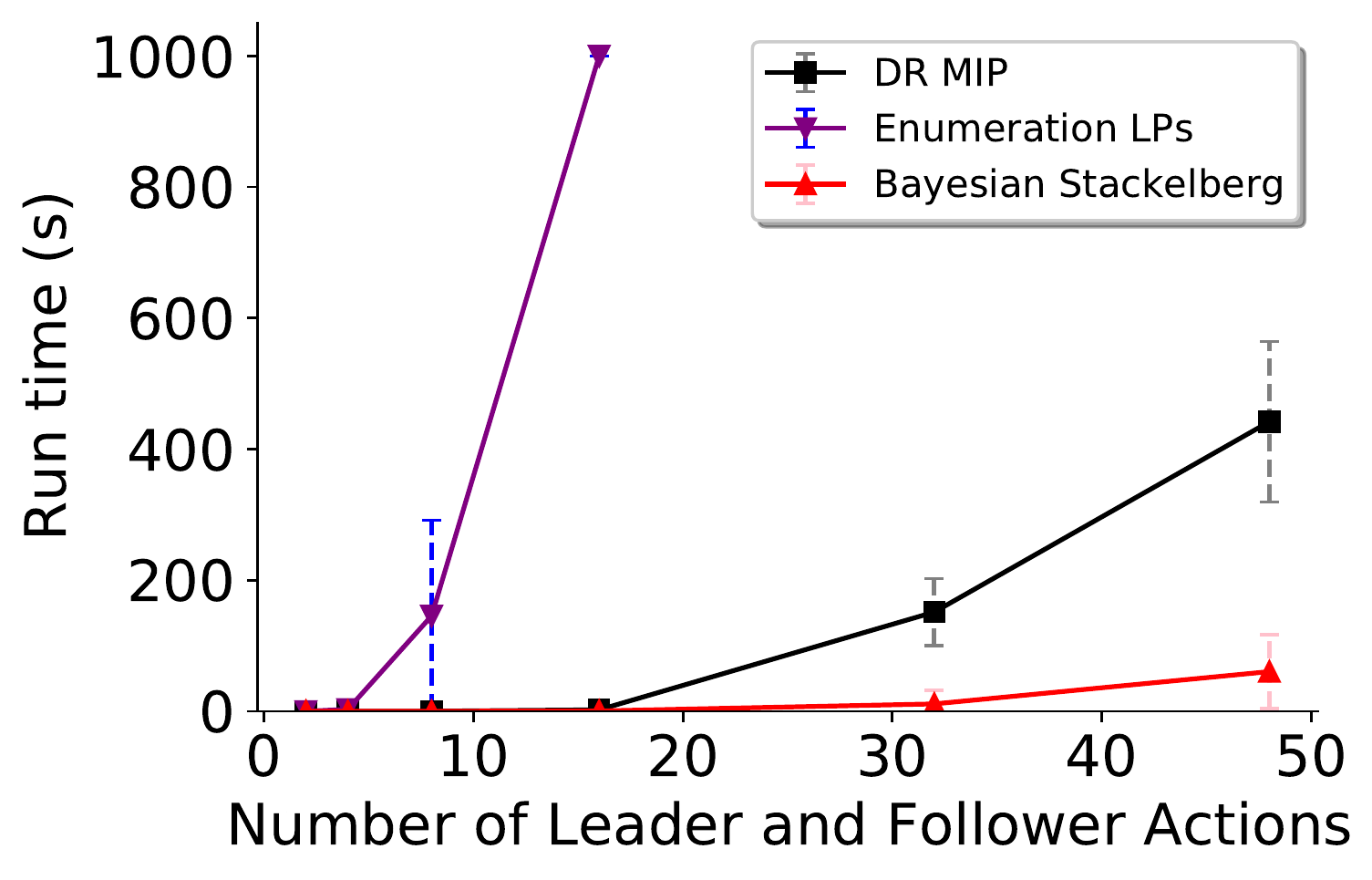}
  \caption{}
  \label{fig:1a}
\end{subfigure} 
\begin{subfigure}{.45\textwidth}
\centering
   \includegraphics[width=\textwidth]{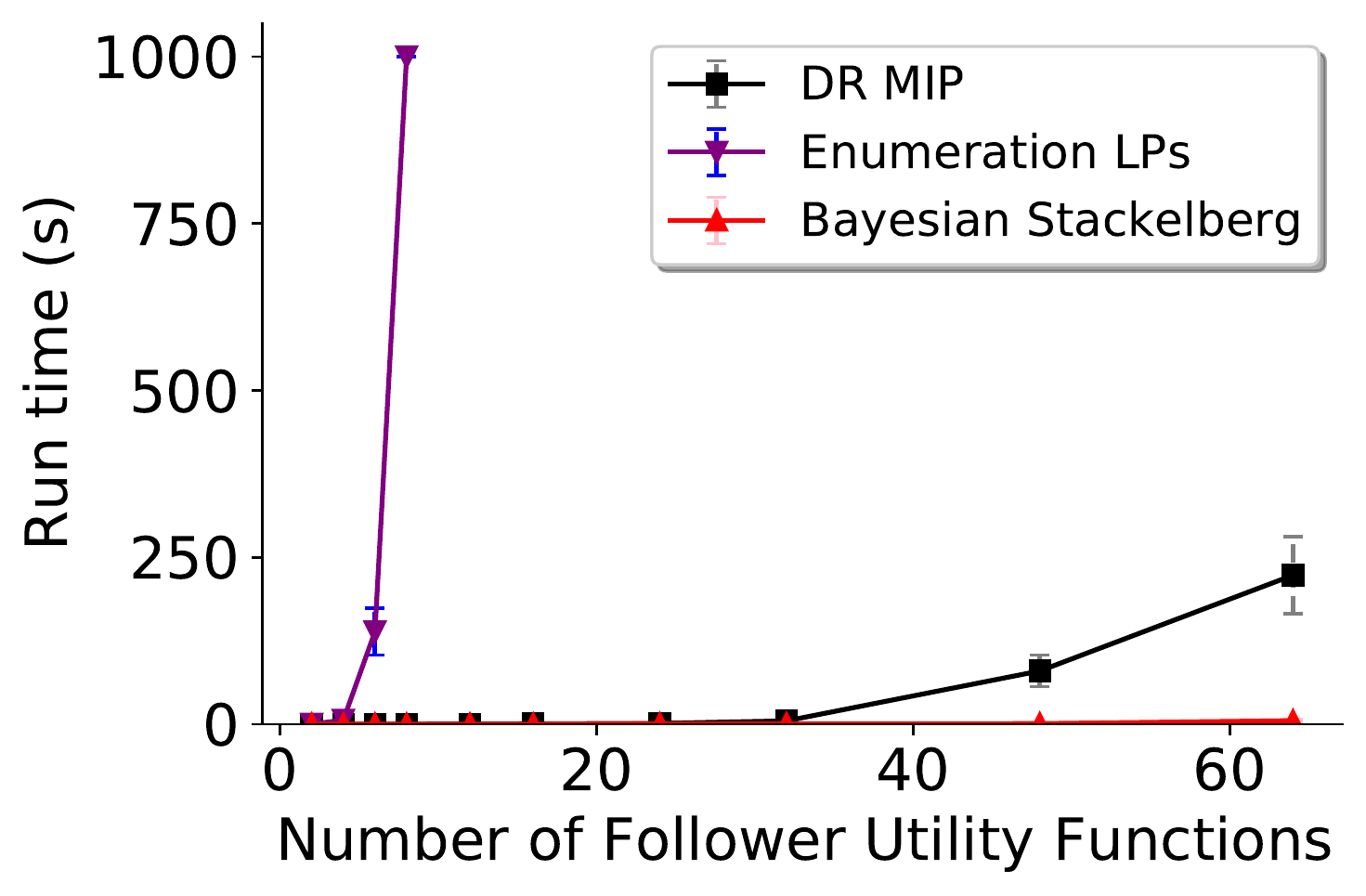}
  \caption{}
  \label{fig:1b}
\end{subfigure} \\
\begin{subfigure}{.5\textwidth}
\centering
   \includegraphics[width=\textwidth]{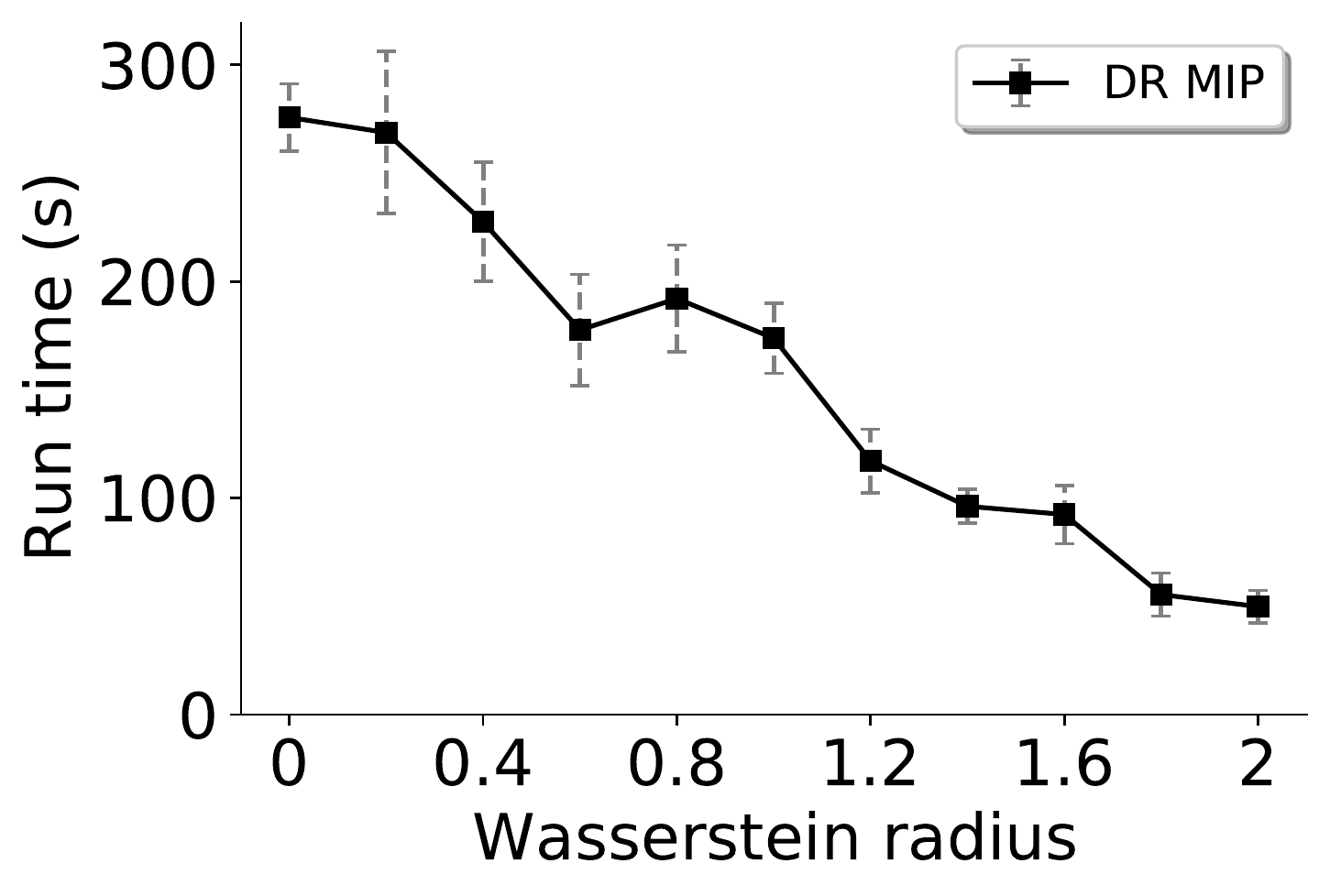}
  \caption{}
  \label{fig:1c}
\end{subfigure}
\label{fig:1}
\caption{Performance of the DR MIP (\ref{mathprog:appendix}) and baselines on the Cournot Duopoly game, averaged over $10$ simulations with standard deviations displayed. \textbf{(a)} Runtime vs number of leader actions ($n$) and follower actions ($m$) with $k=4$. \textbf{(b)} Runtime vs number of follower utility functions ($k$) with $n=m=4$. \textbf{(c)} Runtime vs Wasserstein radius ($\theta$) with $n=m=10$ and $k=12$.}
\end{figure*} \noindent

In Figure \ref{fig:1a}, for $4$ possible follower utility functions, we can compute DRSSE within the threshold for more than $50$ leader and follower actions. The number of follower actions $m$ is crucial to the scalability, affecting the number of integer variables and size of the constraints of the DR MIP (\ref{mathprog:appendix}). The enumeration approach hits the threshold very quickly ($n<20$) and the Bayesian Stackelberg (\ref{mathprog: Bayesian appendix}) runs very fast compared to DRSSE, showing the computational cost of including robustness.

In Figure \ref{fig:1b}, for $4$ leader and follower actions, we can compute DRSSE within the threshold for more than $60$ follower utility functions.  The enumeration approach hits the threshold very quickly ($n<10$) 
since we solve an exponential $m^k$ LPs (one for each $z$). The Bayesian Stackelberg (\ref{mathprog: Bayesian appendix}) runs very fast compared to DRSSE here as well. The number of follower utilities has a moderate impact on scalability of the DR MIP (\ref{mathprog:appendix}).

In Figure \ref{fig:1c}, for $10$ leader and follower actions, and $10$ follower utilities, we vary the Wasserstein radius $\theta$ from $0$ to $2$. The DRSSE problem gets easier to solve with an increase in this parameter, as some small set of utility functions dominate eventually.

\subsection{Results on Synthetic Data}
We present the experimental performance based on a synthetic data set for the utilities: the leader utility and all follower utilities are iid random matrices in $[0,1]$. 

\begin{figure*}[!ht]
\begin{subfigure}{.45\textwidth}
\centering
  \includegraphics[width=\textwidth]{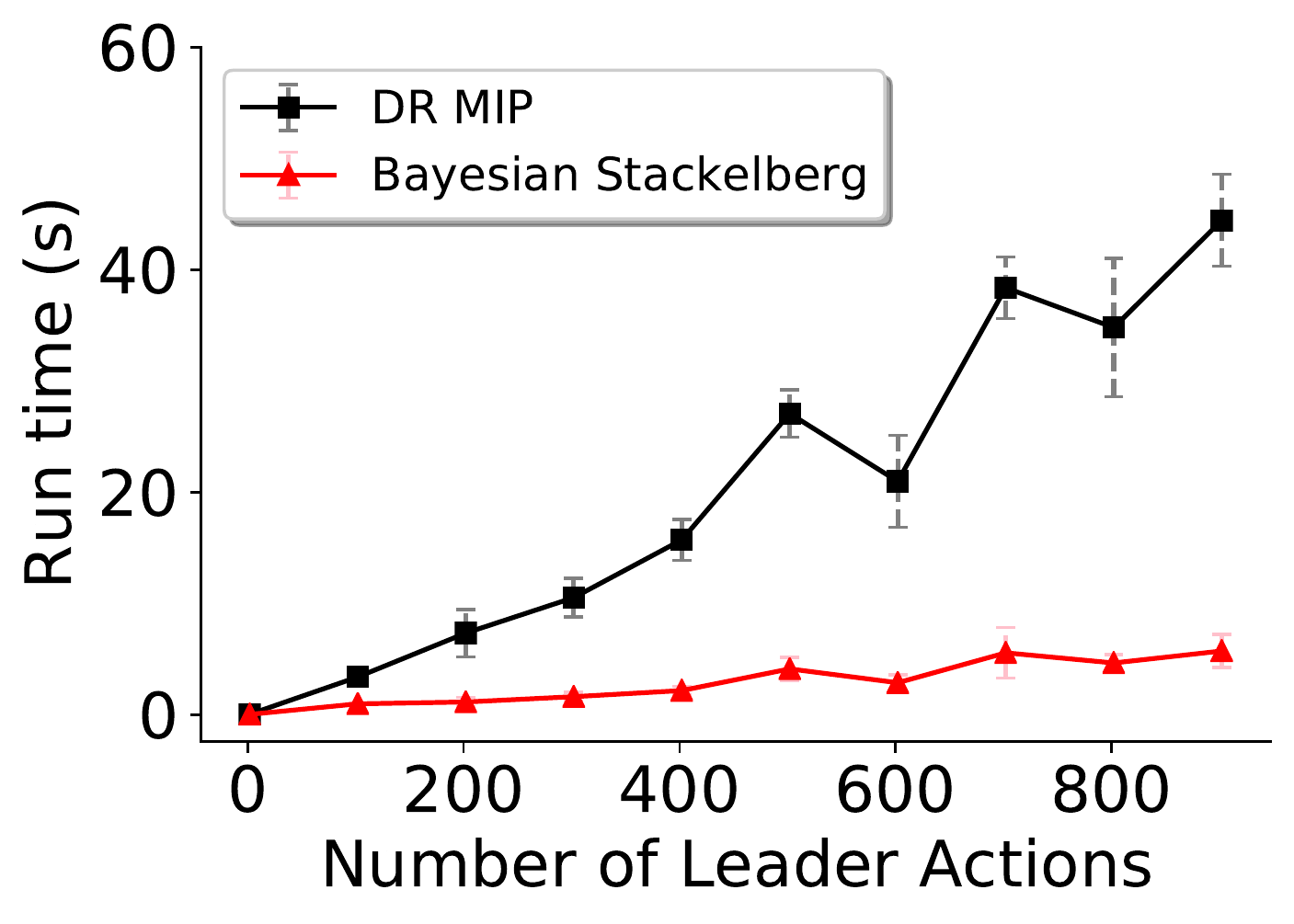}
  \caption{}
  \label{fig:3a}
\end{subfigure} 
\begin{subfigure}{.45\textwidth}
\centering
   \includegraphics[width=\textwidth]{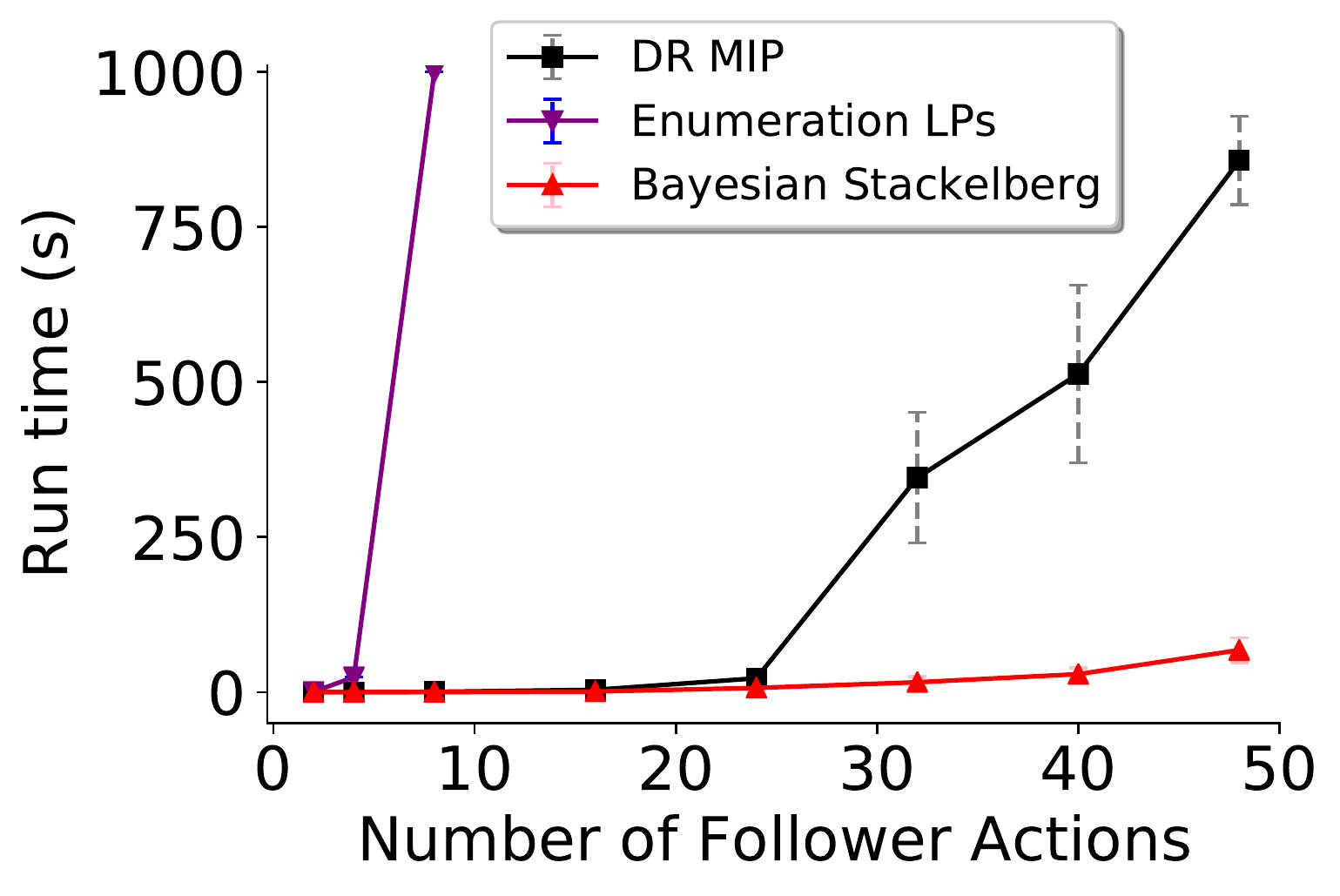}
  \caption{}
  \label{fig:3b}
\end{subfigure} \\
\begin{subfigure}{.45\textwidth}
\centering
   \includegraphics[width=\textwidth]{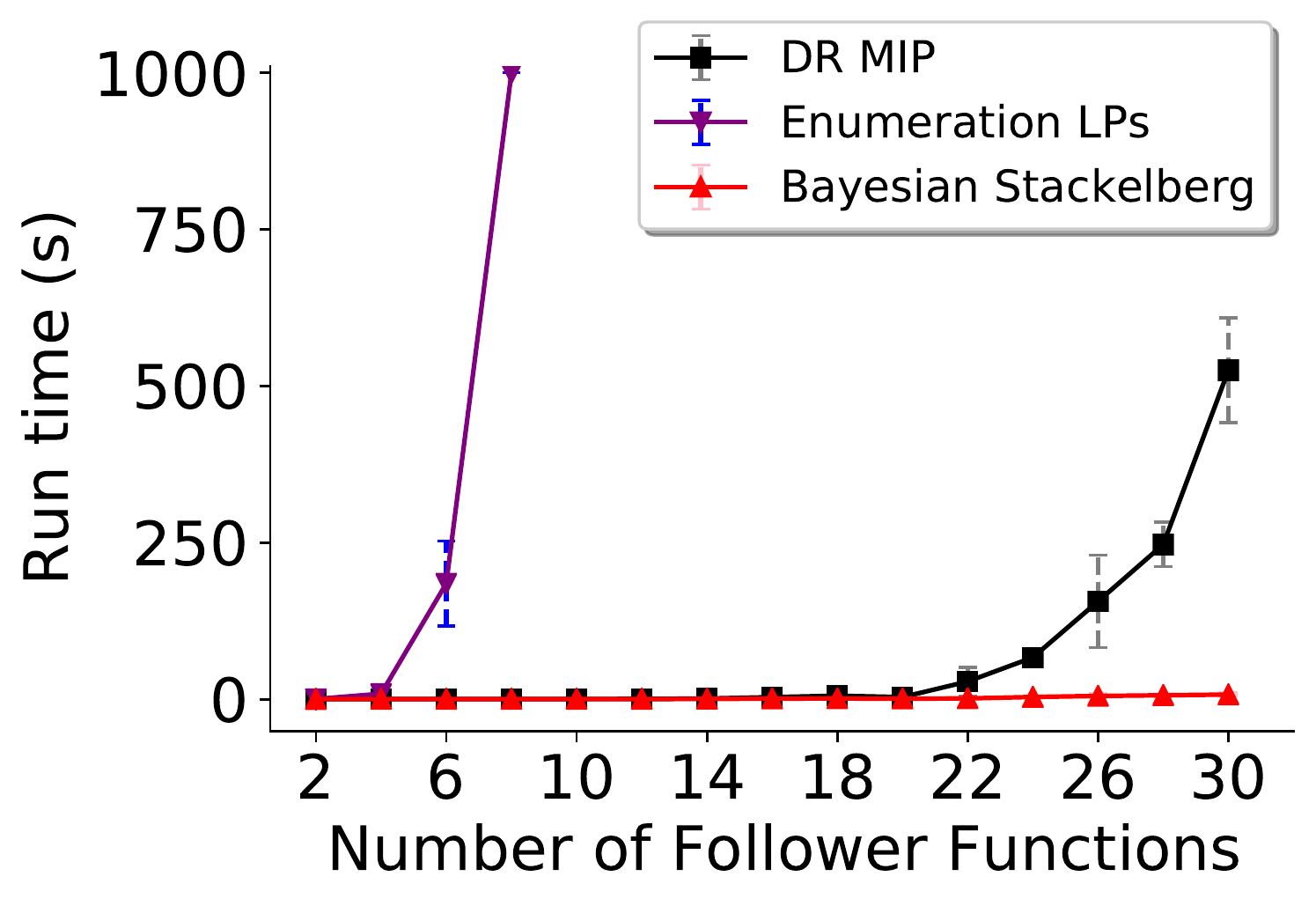}
  \caption{}
  \label{fig:3c}
\end{subfigure}
\begin{subfigure}{.45\textwidth}
\centering
   \includegraphics[width=\textwidth]{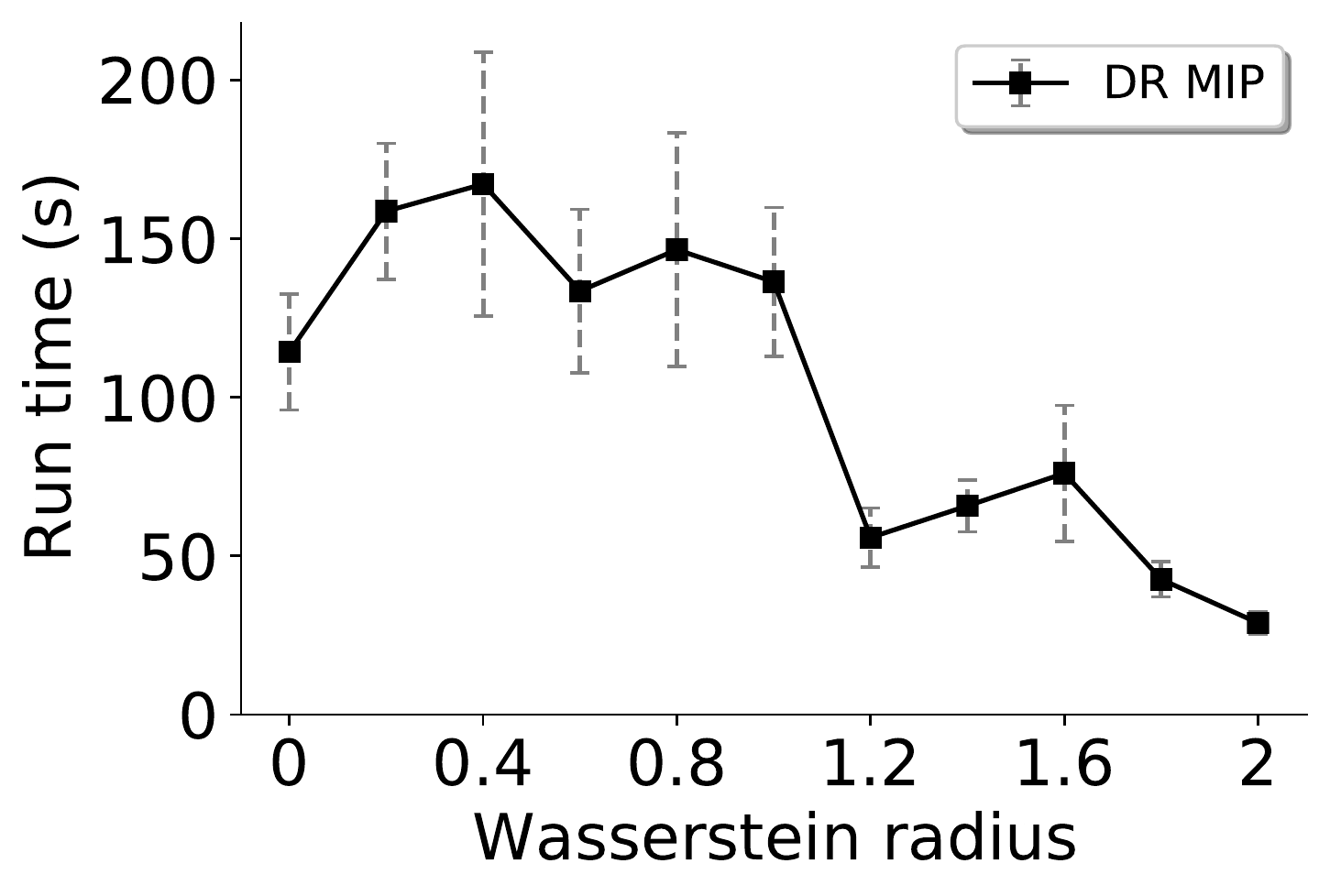}
  \caption{}
  \label{fig:3d}
\end{subfigure}
\label{fig:3}
\caption{Performance of the DR MIP (\ref{mathprog:appendix}) and baselines on the synthetic data set, averaged over $10$ simulations with standard deviations displayed. \textbf{(a)} Runtime vs number of leader actions ($n$) with $m=12$ and $k=4$. \textbf{(b)} Runtime vs number of follower actions ($m$) with $n=50$ and $k=4$. \textbf{(c)} Runtime vs number of follower functions ($k$) with $n=8$ and $k=4$. \textbf{(d)} Runtime vs Wasserstein radius ($\theta$) with $n=m=10$ and $k=12$. }
\end{figure*}

In Figure \ref{fig:3a}, for $m=12$ and $k=4$, we can compute DRSSE within a minute for more than $900$ leader actions. Unsurprisingly, the number of leader actions $n$ is not crucial to the scalability, since these are reflected in continuous variables and not impacting the size of the constraints. The enumeration approach hits the threshold almost immediately  due to the size of action spaces and hence is not plotted. The Bayesian Stackelberg MIP (\ref{mathprog: Bayesian appendix}) runs fast (less than $10$ seconds) compared to DRSSE illustrating the computational cost of including robustness. 

In Figure \ref{fig:3b}, for $n=50$, and $k=4$, we can compute DRSSE within the threshold for more than $50$ follower actions. The enumeration approach hits the threshold very quickly ($m<10$)  since we solve an exponential $m^k$ LPs (one for each $z$). The Bayesian Stackelberg MIP runs very fast compared to DRSSE here as well. The number of follower actions again has a moderate impact on scalability of the DR MIP (\ref{mathprog:appendix}). 

In Figure \ref{fig:3c}, for $n=8$ and $k=4$, we can compute DRSSE within the threshold for more than $30$ follower utility functions. The enumeration approach hits the threshold very quickly ($k<8$) and the Bayesian Stackelberg MIP runs very fast compared to DRSSE here as well. The number of follower utility functions has an exponential impact on the scalability of the DR MIP (\ref{mathprog:appendix}), with $k$ impacting the size of integer variables and integer constraints. 

In Figure \ref{fig:3d}, for $n=m=10$, and $k=12$, we vary the Wasserstein radius $\theta$ from $0$ to $2$. The DRSSE problem gets easier to solve with an increase in this parameter, similar to the Cournot game, as some small set of utility functions dominate eventually.

\end{document}